\documentclass[fleqn,usenatbib]{mnras}

\usepackage{newtxtext,newtxmath}
\usepackage[T1]{fontenc}
\usepackage{ae,aecompl}
\usepackage{url}
\usepackage{graphicx}
\usepackage{xspace}
\usepackage{amsfonts}
\usepackage{pifont}
\usepackage{textcomp}
\usepackage{float}
\usepackage{subfig}
\usepackage{multirow}

\title[The Rotational Phase of Stellar Flares on M dwarfs]{Investigating the Rotational Phase of Stellar Flares on M dwarfs Using K2 Short Cadence Data}

\author[L. Doyle et al.]{
L. Doyle,$^{1,2}$\thanks{E-mail: lauren.doyle@armagh.ac.uk}
G. Ramsay$^{1}$,
J. G. Doyle$^{1}$,
K. Wu$^{3}$,
E. Scullion$^{2}$                                                                                                                                                                                                        
\\
$^{1}$Armagh Observatory and Planetarium, College Hill, Armagh, BT61 9DG\\
$^{2}$Mathematics, Physics and Electrical Engineering, Northumbria University, Newcastle upon Tyne, NE1 8ST\\
$^{3}$Mullard Space Science Laboratory, UCL, Holmbury St. Mary, Dorking, Surrey, RH5 6NT}

\date{Accepted 2018 July 19. Received 2018 July 19; in original form 2017 November 21.}

\pubyear{2017}

\begin{document}

\outer\def\gtae {$\buildrel {\lower3pt\hbox{$>$}} \over 
{\lower2pt\hbox{$\sim$}} $}
\outer\def\ltae {$\buildrel {\lower3pt\hbox{$<$}} \over 
{\lower2pt\hbox{$\sim$}} $}
\newcommand{\Msun}{$M_{\odot}$}
\newcommand{\lsun}{$L_{\odot}$}
\newcommand{\Rsun}{$R_{\odot}$}
\newcommand{\solar}{${\odot}$}
\newcommand{\kep}{\sl Kepler}
\newcommand{\ktwo}{\sl K2}
\newcommand{\swift}{\it Swift}
\newcommand{\Porb}{P_{\rm orb}}
\newcommand{\nuorb}{\nu_{\rm orb}}
\newcommand{\eplus}{\epsilon_+}
\newcommand{\eminus}{\epsilon_-}
\newcommand{\cd}{{\rm\ c\ d^{-1}}}
\newcommand{\MdotL}{\dot M_{\rm L1}}
\newcommand{\Mdot}{$\dot M$}
\newcommand{\Mdotsolar}{\dot{M_{\odot}} yr$^{-1}$}
\newcommand{\Ldisk}{L_{\rm disk}}
\newcommand{\src}{KIC 9202990}
\newcommand{\ergscm} {erg s$^{-1}$ cm$^{-2}$}
\newcommand{\rchi}{$\chi^{2}_{\nu}$}
\newcommand{\chisq}{$\chi^{2}$}
\newcommand{\pcmsq} {cm$^{-2}$}
\maketitle

\begin{abstract}
We present an analysis of K2 short cadence data of 34 M dwarfs which have spectral types in the range M0 - L1. Of these stars, 31 showed flares with a duration between $\sim$10--90 min. Using distances obtained from {\sl Gaia} DR2 parallaxes, we determined the energy of the flares to be in the range $\sim1.2\times10^{29}-6\times10^{34}$ erg. In agreement with previous studies we find rapidly rotating stars tend to show more flares, with evidence for a decline in activity in stars with rotation periods longer than $\sim$10 days. The rotational modulation seen in M dwarf stars is widely considered to result from a starspot which rotates in and out of view. Flux minimum is therefore the rotation phase where we view the main starspot close to the stellar disk center. Surprisingly, having determined the rotational phase of each flare in our study we find none show any preference for rotational phase. We outline three scenarios which could account for this unexpected finding. The relationship between rotation phase and flare rate will be explored further using data from wide surveys such as NGTS and TESS.

\end{abstract}


\section{Introduction}
M dwarfs make up 70\% of the stars in the solar neighbourhood. They are small, cool main sequence stars with temperatures in the range of 2400 - 3800 K and radius between 0.20 - 0.63 \Rsun \citep{gershberg2005solar}. For stars with spectral type later than M4 it is thought their interiors are fully convective \citep{hawley2014kepler}. However, since these stars have no tachocline (a boundary zone between the radiative and convective zones) the star would not be expected to have a significant magnetic field if it was generated in the same manner as in the Sun. Some late M dwarfs {\it do} show strong flaring activity, highlighting a lack of understanding regarding the origins of their activity.

Over the years several models have been developed to account for the presence of a magnetic field in stars which are fully convective. \cite{durney1993generation} propose a dynamo generated by a turbulent velocity field causing the generation of chaotic magnetic fields. \cite{chabrier2006large} explore the possibility of an $\alpha^{2}$ dynamo which can generate large scale magnetic fields. However, this question still remains unresolved and the mechanism for producing a long lived magnetic field in late-type M dwarfs is still unknown.

One method of probing the magnetic field in these late-type stars is to investigate flares using them as a proxy for magnetic activity. In the case of well observed solar flares the correlation between magnetic field in active regions is very clear \citep{fletcher2011observational}. Stellar flares are a phenomena which have been studied for a century. Some of the first detailed optical observations of stellar flares on M dwarfs were made by \citet{bopp1973high} and \citet{gershberg1983characteristics}. Amongst the first X-ray observations of stellar flares were made by \citet{heise1975evidence} using {\sl EXOSAT} who detected an X-ray flare from the M4.5V star YZ CMi. Since then, the physics of stellar flares has been studied by many over the years and in the full energy range from $\gamma$-rays to radio frequencies.

More recently, the {\sl Kepler} mission allowed almost uninterrupted observations of stars lasting many months, or in some cases years \citep{boru2010}. One particularly well studied flare star was GJ 1243 (KIC 9726699) which showed many flares especially when it was observed with a cadence of 1 min (short cadence mode) (\cite{ramsay2013short}, \cite{hawley2014kepler}).  This star has a spectral type of M4 making it an interesting example of a star which is on the cusp of having a fully convective core. \citet{hawley2014kepler} analysed the {\sl Kepler} data of GJ 1243 to identify classical and complex flares, finding correlations between flare energy, amplitude, duration and decay time. However, following the failure of two of {\sl Kepler's} gyroscopes, it was re-purposed as K2  \citep{howell2014k2} observing targets along the ecliptic plane.

M dwarfs can show considerable amplitude variations in their lightcurves which have been explained by the presence of large dominant starspots on the surface moving in and out of view as the star rotates. These amplitude variations in brightness represent one way of determining the stars rotation period and has produced accurate values of rotation periods for thousands of low mass stars observed by Kepler/K2. Using K2 data \citet{stelzer2016rotation} observed 134 M dwarfs and focused on the relation between magnetic activity and stellar rotation.  They found a difference between slow and fast rotating M dwarfs where after a period of approximately 10 days there is an abrupt change in activity. This would suggest a link between activity and rotation but, with the majority of targets being observed in long cadence (30min) mode, short duration flares could not be detected leading to a bias against detecting short duration flares. 

In solar physics the relationship between sunspots and solar flaring activity has been studied for decades and it is generally accepted that these two phenomena are closely related. \cite{guo2014dependence} carried out a statistical study on the dependence of flares in relation to sunspots and  rotational phase in the 22nd and 23rd solar cycles. They found the occurrence of X-class flares was in phase with the solar cycle hence, flares closely follow the same 11 year cycle as sunspots. \cite{maehara2017starspot} investigated the correlation between starspots and superflares on solar-type stars using Kepler observations and identified starspots based on the rotational phase of the brightness minima in the lightcurve showing superflares tend to originate from a larger starspot area. 

Despite the extensive work over the years, one area which has not been investigated in depth is the rotational phase distribution of flares in M dwarfs. If the analogy between the physics of solar and stellar flares holds and these events occur from active regions which typically host spots, then we would expect to see a correlation between starspots and flare occurrence. A small number of stars have been studied to determine whether there is a correlation between stellar rotation phase and number of flares but so far nothing has been found. \cite{ramsay2013short}, \cite{ramsay2015view}, \cite{hawley2014kepler} and \cite{lurie2015kepler} examine the phase distribution of the flares in a small sample of M dwarf stars using Kepler/K2. Each of the stars show flares at all rotational phases despite there being clear rotational modulation present in their lightcurves. There was no evidence for a correlation between rotational phase and number of flares.

This poses the question; do flares show any preference for rotational phase in M dwarfs in general? It is expected that more flares would occur during the minimum of the rotational modulation when the spot is most visible. If this is not the case, then which mechanisms are responsible and what is causing the generation of flares in these active stars? In this paper we look at a sample of M dwarf stars which have been observed in short cadence by K2. By analysing the flare and stellar properties of this group of stars we aim to address some of the questions surrounding the rotational phase distribution of the flares. 

\section{Kepler/K2}
\subsection{Overview}

Kepler was launched in 2009 and studied the same 115 square degree patch of sky just north of the Galactic plane for 4 years, providing extensive photometric data for over 100,000 stars \citep{koch2004overview}. The data obtained by Kepler has revolutionised the study of astrophysics especially in the field of exoplanets where it is responsible for finding the majority through the transit method. In addition, Kepler data has also revolutionised the fields of Asteroseismology, the number of stars with known  rotation periods and interacting binaries. Kepler also provided the means to study stellar flares due to the high precision and length of the lightcurves. Kepler can operate in two observation modes, short cadence (SC) 1 minute exposure and long cadence (LC) 30 minute exposure.

We began a study of late type flare stars towards the end of the Kepler mission. However, this ended with the loss of two of Kepler's reaction wheels. New life was given to the mission when it was re-purposed as K2 which began taking observations in June 2014 where fields are observed along the ecliptic for a duration of $\sim$ 70 - 80 days each \citep{howell2014k2}. This study of late type dwarfs using K2 SC data will allow us to study the flare properties of these stars in greater detail. 

\subsection{Data reparation}
\label{dataprep}

Because of the way K2 is pointed, the center of the stellar point spread function typically moves by $\sim$1 pixel over the course of 6 hrs \citep{vanc16}.  Without applying a photometric correction  the  resulting error on a lightcurve can be considerably higher than for Kepler. For instance, for stars in the 10--11 mag range, Kepler gave an rms of  18 parts per million (ppm). For K2 the uncorrected rms is 170 ppm. Amongst the first to provide readily available corrected data were \citep{vand14} who were able to bring the rms for 10--11 mag stars down to 31 ppm. 

A number of other groups have  developed software which corrects for the instrumental effects which are present in raw K2 data. Not all of these techniques are suitable for SC mode data and some approaches can remove astrophysical effects.  We have used the corrected K2 data using the  EPIC Variability Extraction and Removal for Exoplanet Science Targets ({\tt EVEREST})  pipeline \citep{luger16} in all but one star in our sample.  (For  GJ 1224 observed in Field 9 we  obtained the SC lightcurve from  Andrew Vandenburg).

Each photometric point has raw and corrected fluxes and a `{\tt QUALITY}' flag. This is  of particular importance when trying to find events such as flares which could, in principal, be mistaken for an instrumental effect. The {\tt EVEREST} pipeline keeps the original flag definitions which were used in Kepler, but adds additional `bit values'. For instance, bits `23' and `25' could be detector anomalies but could also be events such as eclipses or flares \citep{luger16,luger17}.  

In searching for photometric  variations longer than the typical duration of flares, such as the stellar rotation period, we were cautious and removed all events which did not conform to {\tt QUALITY=0}. When we searched for flares we removed photometric points which  were identified as bad in the Kepler Archive Manual\footnote{\url{https://archive.stsci.edu/kepler/manuals/archive_manual.pdf}} and  points corresponding to bit values which were clearly identifying times when the spacecraft thruster  was used such as bit value 20 and 21. We kept points which had bit values 23 and 25.

\section{Our K2 late dwarf Sample}
\label{sec:mdwarfs}

To create our sample of stars we took all the sources observed in short cadence with K2 in Fields 1--9 (observations made between May 2014 and Jul 2016) and cross referenced them with {\tt SIMBAD}\footnote{\url{http://simbad.u-strasbg.fr/simbad}}, removing all stars which were not of spectral type M0 or later. In addition, those stars classed as BY Dra stars were removed from our sample since many of them are in binaries or triples and are therefore, not conducive to investigating stellar activity on single stars before and after the M4 spectral sub-type. The remaining stars were then cross referenced with the EPIC catalogue and those which showed characteristics of a giant (e.g. have a radii >1\Rsun) were also removed. Stars which were too faint to show a clear detection in the K2 thumb print image were also removed from our sample.

This left us with a sample of late dwarf stars observed with K2 in short cadence mode, consisting of 33 M dwarfs and one L dwarf. Of this sample of stars, 32 percent were classed as known flare stars in the {\tt SIMBAD} catalogue. They range in spectral type and mass from M0 to L1 and 0.58$M_{\odot}$ to 0.08$M_{\odot}$. Each target has been observed for $\sim$70 -- 80 days producing a near continuous lightcurve over this period.  The significance of short cadence data is it allows flares with a duration of a few minutes to be detected, giving a more comprehensive and robust overview of stellar activity.  In addition, the wide range of spectral types provides a broader insight into how magnetic activity can vary in M dwarfs as a whole. The properties of the stars in our sample, including spectral type, previously known rotation period, mass, distance and magnitude are shown in Table \ref{mdwarfs_lit}. 

\begin{table*}
\caption{The properties of the M stars which are included in our survey where all data comes from the K2 EPIC Catalog \citep{Huber2016},
with the exception of the following. References for spectra: [1] \citep{Hawley1996}, [2] \citep{Reid2008}, [3] \citep{Reid2004}; [4] \citep{Lepine2011}; [5] \citep{Kraus2007}, [6] \citep{Schmidt2010}, [7] \citep{Gray2003}; [8] \citep{Davison2015}; [9] \citep{lepine2013spectroscopic}; [10] \citep{kirkpatrick1991standard}; [11] \citep{faherty2009proper}; [12] \citep{pesch1968late}; [13] \citep{alonso2015carmenes}; [14] \citep{stephenson1986dwarf}; [15] \citep{cruz2002meeting}; [16] \citep{shkolnik2009identifying}.}
\begin{center}
\resizebox{\textwidth}{!}{%
\footnotesize
\begin{tabular}{lclccrrcccccccc}
\hline 
Name                         & EPIC       & K2     & RA            & DEC            & SpT             & P         & M     & R       & T$_{eff}$ & log g \\
                             & ID         & Field  & (J2000)       & (J2000)        &                 & d         & \Msun & \Rsun   & (K)       & (cgs) \\
\hline 
LHS 2420                     & 201611969  & 1      & 11:31:32.845  & +02:13:42.86   & M2.5V [9]       &           & 0.319 & 0.301   & 3740      & 4.977 \\
LP 804-27                    & 205204563  & 2      & 16:12:41.781  & --18:52:31.83  & M3V [7]         &           & 0.445 & 0.391   & 3930      & 4.86  \\
GJ 3954                      & 205467732  & 2      & 16:26:48.160  & --17:23:33.6   & M4.5V           &           & 0.304 & 0.286   & 3772      & 4.995 \\
IL Aqr                       & 206019387  & 3      & 22:53:16.7    & --14:15:49.3   & M4V [1]         & 95        & 0.21  & 0.22    &  3472     & 5.07  \\
LP 760-3                     & 206050032  & 3      & 22:28:54.401  & --13:25:17.86  & M6.5V [10]      &           & 0.09  & 0.113   & 2617      & 5.271 \\
2MASSI J2214-1319            & 206053352  & 3      &  22:14:50.707 & --13:19:59.080 & M7.5 [11]       &           & 0.078 & 0.098   & 2211      & 5.349 \\
Wolf 1561 A                  & 206262336  & 3      & 22:17:18.9    & --08:48:12.5   & M4V+M5V [1]     &           & 0.22  & 0.23    &  3495     & 5.06  \\
HG 7-26                      & 210317378  & 4      &3:52:34.340    & +11:15:38.807  & M1 [12]         & 0.192     & 0.396 & 0.358   & 3788      & 4.917 \\
NLTT 12593                   & 210434433  & 4      & 4:07:54.80    & +14:13:00.7    & M2.5V [13]      & 1.073     & 0.424 & 0.382   & 3844      & 4.896 \\
G 6-33                       & 210460280  & 4      & 3:45:54.83    & +14:42:52.1    & M1.5 [9]        &           & 0.511 & 0.454   & 4028      & 4.821 \\
LP 415-363                   & 210489654  & 4      & 4:20:47.988   & +15:14:09.073  & M4V [13]        & 82.6      & 0.391 & 0.351   & 3751      & 4.917 \\
MCC 428                      & 210579749  & 4      & 3:43:45.247   & +16:40:02.166  & M0V [14]        &           & 0.504 & 0.449   & 3898      & 4.836 \\
GJ 3225                      & 210758829  & 4      & 03:26:45.0    & +19:14:40.1    & M4.5V [1]       & 0.454     & 0.13  & 0.15    &  3027     & 5.19  \\
2MASS J0326+1919             & 210764183  & 4      & 03:26:44.5    & +19:19:31.0    & M8.5V [2]       &           & 0.08  & 0.10    &  2192     & 5.34  \\
LP 414-108                   & 210811310  & 4      & 04:10:38.1    & +20:02:23.5    & M0.5V+M0.5V [3] &           & 0.43  & 0.37    &  3971     & 4.89  \\
LP 357-206                   & 210894955  & 4      & 3:55:36.90    & +21:18:48.30   & M5 [15]         & 7.916     & 0.240 & 0.248   & 3550      & 5.014 \\
2MASSJ0335+2342              & 211046195  & 4      & 3:35:02.087   & +23:42:35.61   & M8.5V [16]      & 0.472     & 0.084 & 0.106   & 2432      & 5.310 \\
LT Tau                       & 211069418  & 4      & 03:42:56.5    & +24:04:58.1    & M3.5V [4]       &           & 0.32  & 0.30    &  3675     & 4.97  \\
V497 Tau                     & 211077349  & 4      & 03:42:02.9    & +24:12:36.3    & M3V [4]         &           & 0.44  & 0.40    &  3876     & 4.87  \\
V692 Tau                     & 211082433  & 4      & 03:56:30.4    & +24:17:18.8    & M4-5V [4]       &           & 0.26  & 0.26    &  3567     & 5.03  \\
V631 Tau                     & 211112686  & 4      & 03:44:24.8    & +24:46:06.3    & M1V [4]         & 3.27      & 0.37  & 0.34    &  3700     & 4.93  \\
V* MY Tau                    & 211117230  & 4      & 3:44:27.293   & +24:50:38.26   & M0 [16]         & 0.4       & 0.427 & 0.381   & 3811      & 4.894 \\
2MASS J0831+1025             & 211329075  & 5      & 08:31:56.0    & +10:25:41.7    & M9V [2]         &           & 0.08  & 0.10    &  2209     & 5.34  \\
GJ 3508                      & 211642294  & 5      & 8:37:07.961   & +15:07:45.5257 & M3V [9]         &           & 0.427 & 0.385   & 3861      & 4.892 \\
LP 426-35                    & 211945363  & 5      & 08:57:15.4    & +19:24:17.7    & M5V [5]         &           & 0.35  & 0.32    &  3738     & 4.96  \\
2MASS J0909+1940             & 211963497  & 5      & 09:09:48.2    & +19:40:42.9    & L1 [6]          &           & 0.08  & 0.10    &  2098     & 5.36  \\
AX Cnc                       & 211970427  & 5      & 08:39:09.9    & +19:46:58.9    & M2 [5]          & 4.854     & 0.45  & 0.41    &  3881     & 4.87  \\
2MASS J0831+2024             & 212009427  & 5      & 08:31:29.9    & +20:24:37.5    & M0V [5]         &  1.55     & 0.58  & 0.51    &  4116     & 4.78  \\
2MASS J0839+2044             & 212029094  & 5      & 08:39:18.1    & +20:44:21.3    & M1V [5]         &           & 0.34  & 0.32    &  3785     & 4.96  \\
L 762-51                     & 212285603  & 6      & 13:45:50.7    & --17:58:05.6   & M3.5V [1]       &           & 0.31  & 0.30    &  3646     & 4.99  \\
LP 737-14                    & 212518629  & 6      & 13:16:45.47   & --12:20:20.4   & M3.5V [13]      &           & 0.281 & 0.273   &  3577     & 5.005 \\
BD-05 3740                   & 212776174  & 6      & 13:38:58.7    & --06:14:12.5   & M0.5V [7]       &           & 0.45  & 0.40    &  3847     & 4.88  \\
2MASSI J1332-0441            & 212826600  & 6      & 13:32:24.427  & --4:41:12.690  & M7.5 [6]        &           & 0.079 & 0.101   & 2264      & 5.328 \\
GJ 1224                      & 228162462  & 9      & 18:07:32.927  & --15:57:46.46  & M4V [8]         &           & 0.14  &         &           &       \\
\hline
\end{tabular}
}
\label{mdwarfs_lit}

\end{center}
\end{table*}

\section{Rotation Period}
\label{rotation}

From our sample of 34 stars, only five have rotation periods previously recorded in the literature, see Table \ref{mdwarfs_lit}. The two most important factors in determining the level of activity in a low mass star are age and rotation period. The rotation period can be determined from low mass stars if they have starspots whose rotation can produce a change in brightness as the starspots are expected to be darker than the photosphere (in the same way as sunspots). More recently \cite{karoff2018influence} compared the characteristics of the solar analog star HD173701 with the Sun and found that metallicity could also be important factor in setting activity levels.  How this results in a stronger dynamo is still unclear. 

We first set out to determine or constrain the rotation period from the K2 data. The method used to determine the rotation period is initially through the Lomb-Scargle (LS) periodogram. We define $\phi$=0.0 as the rotational phase where the flux is at a minimum and is first obtained by eye. For stars with a modulation period longer than 10 d we derive the rotation period, $P_{rot}$, from the LS periodogram and fold the lightcurve to obtain a phase folded lightcurve. For stars where $P_{rot}<10$ d, we are able to refine $P_{rot}$, and the time $T_{0}$, which defines the first minimum, by phasing and folding sections of the lightcurve taken from the start, middle and end. This iterative procedure allows allows $P_{rot}$ and $T_{0}$ to be determined more accurately than the LS periodogram alone. This gives us a mean folded lightcurve for our sample and is used during the analysis of flaring activity on the targets. The uncertainty on the rotation period is estimated by determining the Full Width at Half Max (FWHM) of the corresponding peak on the power spectrum. 

We show the stellar rotation period derived for our sample using K2 data in Table \ref{Mdwarf_period}: they  range from 0.21 days to greater than 70 days. There are six stars in our sample where the amplitude modulation was incomplete and therefore it was only possible to say the rotation period is greater than the observation length, i.e. greater than 70 days. A further two stars had no apparent modulation in their lightcurve (possibly because they lack a large starspot). We find no correlation between rotation period and spectral type. For instance, stars with a rotation period less than 1 day have spectral types in the range M0--M8.5.  Other examples of rapid rotation in late type active stars include the M7 dwarf 2MASS J0335+23 \citep{gizis2017} which has a very rapid rotation period of 0.22 d, which is presumably due to its young age (24 Myr). \cite{gizis2017} find 22 flares in the K2 lightcurve of 2MASS J0335+23 showing it is active.  

Taking a sample of $\sim$12,000 main sequence stars observed using Kepler, \cite{nielsen2013rotation} found M dwarfs had a median period of 15.4 d, although with a considerable spread in the overall distribution. Out of all of our stars only a handful show rotation periods which are not in agreement with the literature (compare column 7 in Table \ref{mdwarfs_lit} with column 2 in Table \ref{Mdwarf_period}). For some objects, the difference can be substantial, for example, HG 7-26 (EPIC 210317378) had a reported period of 0.192 days \citep{newton2016rotation} compared to 24.5 days as derived from the K2 data, (although the source is also classed as a `Non detection or undetermined detection' in the catalogue of \cite{newton2016rotation}). This underlines the need for a continuous sequence of photometric measurements, such as provided by K2, to reliably determine rotation periods.

\section{Flare Identification}
\label{fbeye}

The flare identification process was completed using Flares By EYE ({\tt FBEYE}), a suite of IDL programs created by J.R.A. Davenport \citep{davenport2014kepler}. {\tt FBEYE} allows users to manually classify flares present in the lightcurve via an interactive display. We are therefore able to determine for each flare the peak time, start time, end time, flux peak and equivalent duration. The lightcurves which we used for this process were complete, meaning all photometric points were used regardless of their quality flag due to potential flaring events having quality flags which are not zero. The {\tt EVEREST} quality flags and {\tt FBEYE} flare list were then compared directly to assess the likelihood of the flare being a real event. 
Any photometric point with quality flags which were a result of thruster firing or known instrumental effects were removed from further analysis. Any point which had a flag of {\tt EVEREST} bits 23 and 25 (which may have been due to cosmic rays or a real stellar flares) were kept. Events which consisted of only one photometric point were removed and events which did not have profiles consistent with being a likely stellar flare (i.e. sharp rise and exponential decay) were also removed. 

\begin{table*}
\caption{For the stars in our survey which show flaring activity we indicate their observed rotation period; the number of flares together with their duration, amplitude and energy. Since the length of each observation differs by a small amount, we have included a normalised flare number which is the number of flares expected on each star if the observation duration was 78.3 d.}

   \begin{center}
   \label{Mdwarf_period}

	\begin{tabular}{|c|c|c|c|c|c|c|}
    \hline 
    EPIC ID     & Rotation Period ($P_{rot}$) & No. of flares & Normalised Flare No. & Duration Range   & Amplitude Range   & $log(E_{Kp})$    \\ 
                & days                        &               &                      & minutes          & Flux --           & ergs            \\ %
                &                             &               &                      &                  &                                     \\ 
    \hline
    201611969   & > 70                        & 15            &  14.33               & 11.8 -- 35.3     & 0.0009 -- 0.0095  & 31.17 -- 32.18  \\ 
    205204563   & 42:                         & 5             &  4.72                & 8.83 -- 20.6     & 0.0003 -- 0.0030  & 29.12 -- 31.05  \\ 
    205467732   & 1.321 $\pm$ 0.021           & 221           &  205.35              & 8.83 -- 91.2     & 0.0003 - 1.3694   & 30.60 -- 33.49  \\ 
    206019387   & > 70                        & 47            &  39.27               & 10.8 -- 81.4     & 0.0004 -- 0.0151  & 30.01 -- 31.77  \\ 
    206050032   & > 70                        & 17            &  14.21               & 10.8 -- 32.4     & 0.0274 -- 0.6937  & 31.25 -- 32.52  \\ 
    206053352   & --                          & 8             &  6.78                & 11.8 -- 30.4     & 0.2254 -- 6.7031  & 31.19 -- 32.27  \\ 
    206262336   & 9.6 $\pm$ 1.2               & 237           &  194.73              & 8.78 -- 46.1     & 0.0018 -- 1.2462  & 29.45 -- 33.06  \\ 
    210317378   & 24.5:                       & 7             &  6.06                & 12.8 -- 21.6     & 0.0069 -- 0.0582  & 31.39 -- 32.28  \\ 
    210434433   & 47:                         & 4             &  3.49                & 14.7 -- 24.5     & 0.0041 -- 0.0207  & 31.02 -- 32.11  \\ 
    210460280   & 45:                         & 1             &  0.87                & 29.42            & 0.0024            & 31.47           \\ 
    210489654   & > 80                        & 24            &  20.90               & 10.8 -- 39.2     & 0.0037 -- 0.0632  & 31.06 -- 32.28  \\ 
    210579749   & 23:                         & 4             &  3.43                & 12.8 -- 42.2     & 0.0014 -- 0.0027  & 31.37 -- 31.95  \\ 
    210758829   & 0.4539 $\pm$ 0.0027         & 197           &  170.68              & 8.78 -- 106      & 0.0023 -- 4.3192  & 30.09 -- 33.46  \\ 
    210764183   & 0.966 $\pm$ 0.011           & 10            &  8.77                & 8.78 -- 45.9     & 0.2365 -- 5.1922  & 31.95 --        \\ 
    210811310   & 33:                         & 2             &  1.75                & 10.8             & 0.0013 -- 0.0028  & 30.88 -- 31.16  \\ 
    210894955   & 0.726  $\pm$ 0.006          & 18            &  15.81               & 10.8 -- 30.4     & 0.0455 -- 2.3884  & 30.65 -- 32.58  \\ 
    211046195   & 0.2185 $\pm$ 0.0006         & 16            &  14.02               & 0.98 -- 2888     & 0.0437 -- 4.5665  & 31.15 -- 34.57  \\ 
    211069418   & 0.8177 $\pm$ 0.0086         & 104           &  90.48               & 8.78 -- 63.8     & 0.0101 -- 0.4138  & 31.65 -- 33.81  \\ 
    211077349   & 0.6992 $\pm$ 0.0061         & 64            &  55.75               & 8.78 -- 75.6     & 0.0142 -- 3.8455  & 31.69 -- 34.77  \\ 
    211082433   & 0.3447 $\pm$ 0.0015         & 123           &  107.23              & 8.78 -- 71.6     & 0.0097 -- 0.4683  & 31.17 -- 33.51  \\ 
    211112686   & 0.7639 $\pm$ 0.0073         & 53            &  46.34               & 8.78 -- 94.0     & 0.0103 -- 0.4311  & 31.99 -- 34.24  \\ 
    211117230   & 0.398 $\pm$ 0.002           & 38            &  32.72               & 8.83 -- 76.5     & 0.0045 -- 0.2136  & 32.08 -- 34.28  \\ 
    211642294   & 52:                         & 5             &  4.49                & 20.6 -- 103      & 0.0023 -- 0.0196  & 31.15 -- 34.42  \\ 
    211945363   &  > 70                       & 8             &  7.39                & 10.8 -- 31.2     & 0.0025 -- 0.0231  & 31.19 -- 31.98  \\ 
    211970427   & 4.38 $\pm$ 0.24             & 51            &  47.31               & 8.78 -- 64.8     & 0.0101 -- 0.3572  & 32.04 -- 34.09  \\ 
    212009427   & 1.556 $\pm$ 0.029           & 75            &  69.34               & 8.78 -- 76.5     & 0.0068 -- 0.0967  & 32.14 -- 34.29  \\ 
    212029094   & 20.22 $\pm$ 5.03            & 3             &  2.78                & 14.7 -- 22.5     & 0.0019 -- 0.0087  & 31.19 -- 31.66  \\ 
    212518629   & 80:                         & 1             &  0.96                & 27.46            & 0.0106            & 31.46           \\ 
    212776174   & 18.35 $\pm$ 5.03            & 7             &  6.58                & 10.8 -- 20.6     & 0.0007 -- 0.0038  & 31.21 -- 31.99  \\ 
    212826600   & --                          & 7             &  6.79                & 12.8 -- 41.2     & 0.0015 -- 0.8707  & 32.15 -- 33.31  \\ 
    228162462   & 3.9 $\pm$ 0.2               & 424           &  355.31              & 8.78 -- 165      & 0.0015 -- 0.8707  & 29.32 -- 32.74  \\ 
    \hline
    \end{tabular}
	\vspace{0.1cm}\scriptsize{\begin{flushleft}
\textbf{Notes.} For a handful of sources the apparent modulation period is longer than the observation length meaning only an lower limit to the rotation period could be determined. For stars with no evidence for a modulation no rotation period could be determined.
 \end{flushleft}}
 \end{center}
\end{table*}

The data for each star were analysed in a consistent manner and for each star we show the number of flares, the range in the duration and amplitude of the flares in  Table \ref{Mdwarf_period}. In addition, we have added the normalised flare number which represents the number of flares expected on each star if the observation duration was 78.3 days. The duration of each flare was calculated from the start and stop times and the amplitude represents the flux peak of each flare all which were obtained from the output of {\tt FBEYE}. Sources 2MASS J0831+1025, 2MASS J0909+1940 and L 762-51 did not show any flares and were omitted from Table \ref{Mdwarf_period} -- these sources did also not show complete rotational cycles and so the rotational periods could not be determined.

\begin{figure*}
\centering
\includegraphics[width = 0.95\textwidth]{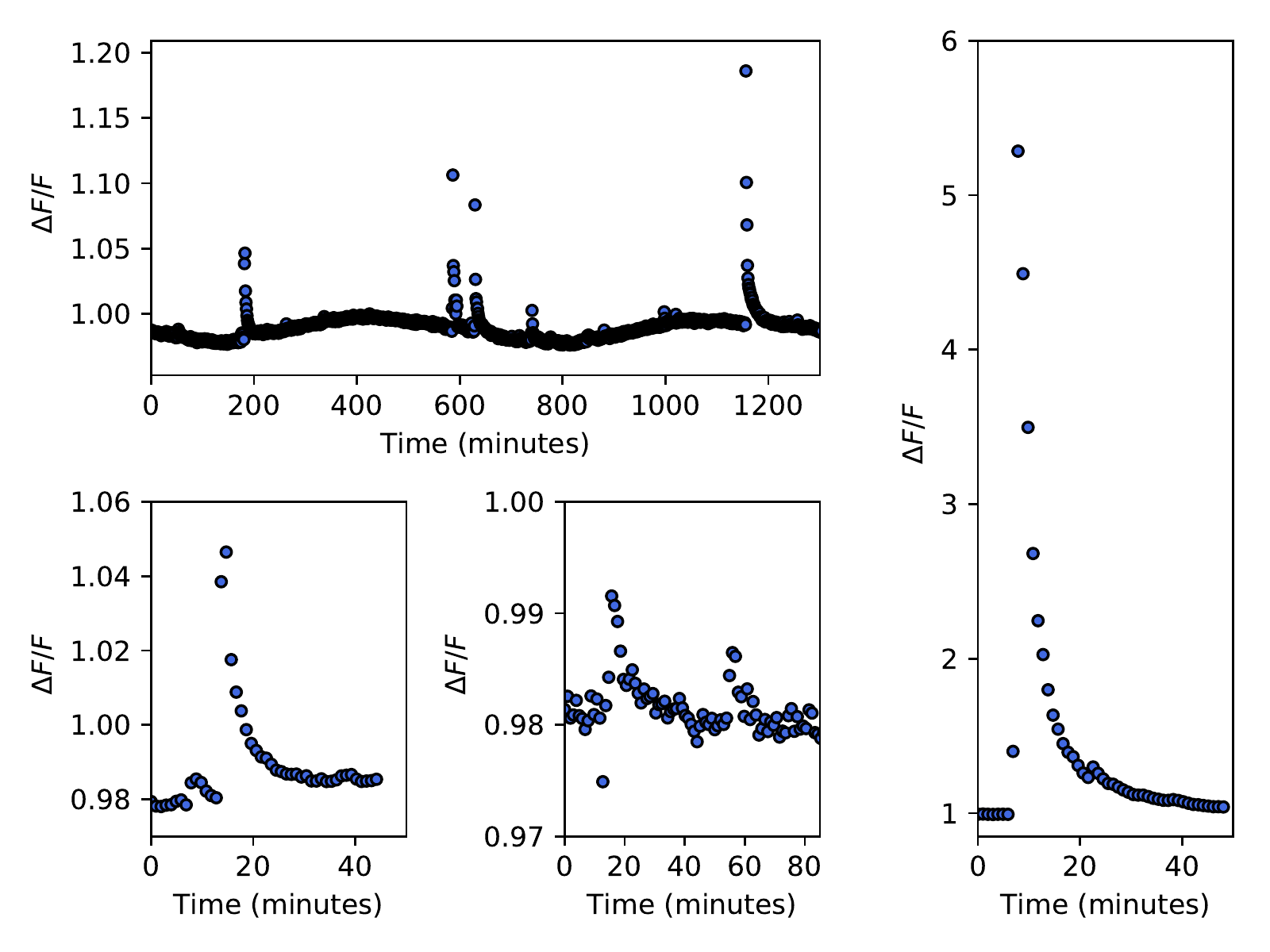}
\caption{A sample of flares of varying magnitude from the flare star GJ 3225 (EPIC 210758829) a M4.5 star with a rotation period of 0.45 days. The top plot shows a small section of the lightcurve of the star demonstrating the frequency of the flares and also the range in magnitude. Far right is the largest flare seen in the K2 lightcurve of this star with a peak normalised flux of 4.32. The bottom two plots show smaller amplitude, short duration flares which dominate the lightcurve. }
\label{flare_image}
\end{figure*}

We show some examples of flares from GJ 3225 (EPIC 210758829) in Figure \ref{flare_image}. Of particular interest is the largest flare from this star with a normalised flux peak of 4.32 (equivalent of 1.6 mag), where a rapid rise (approximately 1 min) and clear slow decay (approximately 10 min) can be seen which, is similar to a classical stellar flare profile.

\begin{figure}
\centering
\includegraphics[width = 0.5\textwidth]{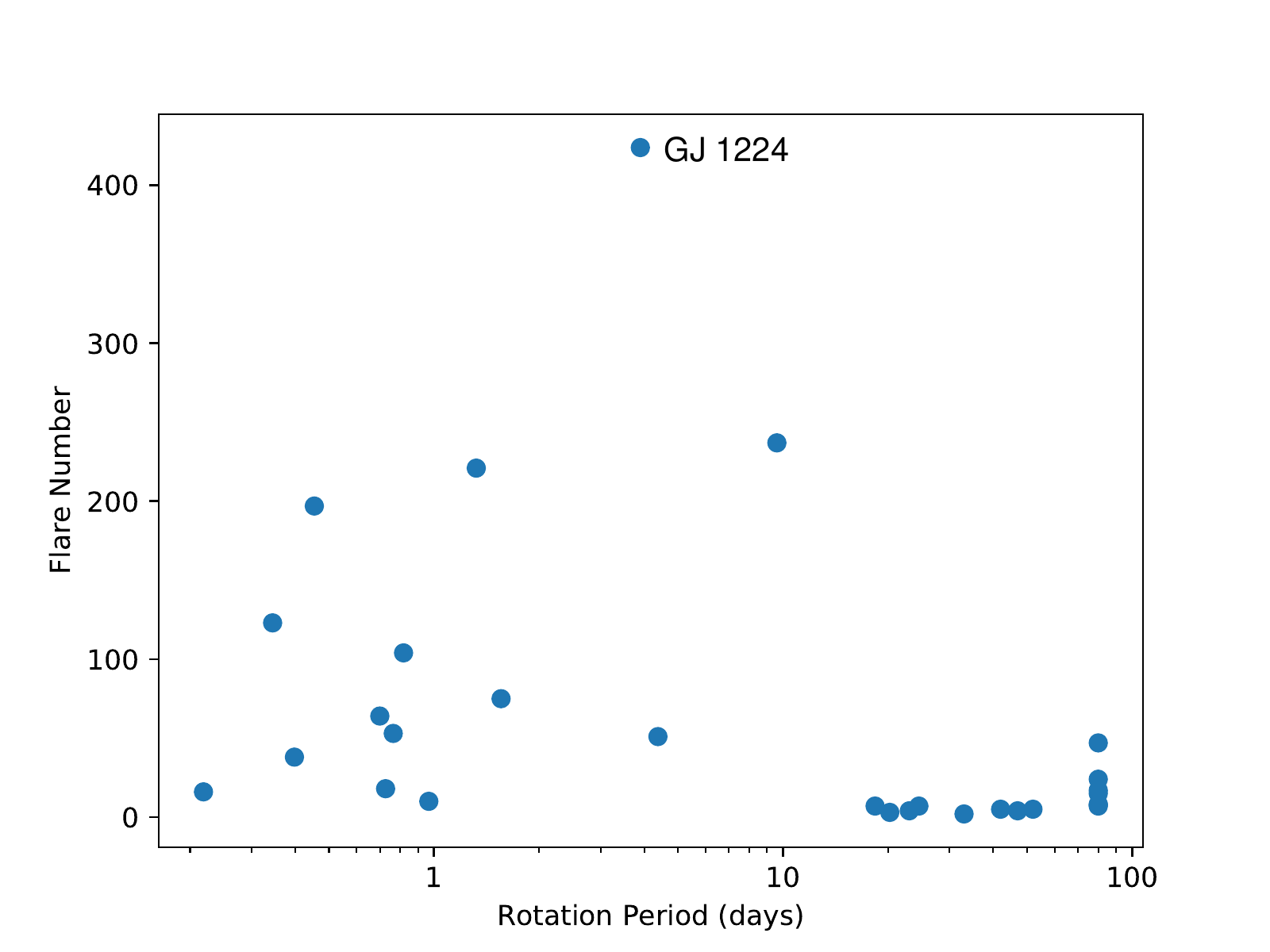}
\caption{The number of flares as a function of the rotation period for all stars showing flaring activity. As other studies have found, stars with rotation periods longer than $\sim$10 d are less active. The most active star is GJ 1224, $P_{rot}=$3.9 d. There are six sources to the very right hand side which represent the stars with rotation periods $>$ 70 d.}
\label{rot_vs_flareno}
\end{figure}

It is expected to find greater flare activity on stars with shorter rotation periods. This is the case with this particular sample of M dwarfs (Figure \ref{rot_vs_flareno}), however, the star with the most flares does not have the fastest rotation. After a rotation period of approximately 10 days there is a drop off in the number of flares seen on the star, which is consistent with the findings of \cite{stelzer2016rotation}. In order to create a complete picture of this sample of M dwarfs the ages of the stars would need to be determined, as although the activity of the stars depends on rotation, this in turn depends on age. 

\section{Flare Energies}

To determine the energy of the stellar flares, the quiescent luminosity of the star, $L_{*}$, must be calculated in the Kepler bandpass. To construct a template spectral energy distribution for each star we used PanStarrs magnitudes g, r, i and z (see Table \ref{Panstarrs_mag}). The PanStarrs magnitudes \citep{chambers2016panstarrs} represent the mean quiescent magnitude of the star calculated over an average of multiple measurements. This ensures the effects of flares or rotation has been removed from the magnitudes. We fitted a polynomial to the Panstarrs data and determined the flux in the Kepler bandpass in the same manner as \cite{kowalski2013time}. The quiescent luminosity is then computed by multiplying the flux by $4\pi d^2$, where the distance ($d$) to the stars are have been determined by inverting the parallax from the {\sl Gaia} Data 2 release \citep{gai16,gai18}, (see Table \ref{Gaia_data}). We also used the Bayesian analysis approach as described by \cite{astra16} and implemented in the {\tt STILTS} suite of software \citep{taylor2006}, and find that, as expected for sources within 200 pc, the distances and errors determined using the inversion and the Bayesian approach are entirely consistent. 

The energy of the flares, $E_{\rm flare}$, can then be determined by the multiplication of the luminosity of the star, $L_{*}$, in erg/s and the equivalent duration, $t$, in seconds. The equivalent duration \citep{gershberg1972some} is defined as the area under the flare lightcurve in units of seconds, which is different to the flare duration, and is obtained through the {\tt FBEYE} suite of programs. {\tt FBEYE} uses a Trapezoidal summation of the area under the flare lightcurve which is converted into seconds. 

A wide range of flare energies are seen in the 31 flaring M dwarfs. The most energetic flare is observed in V497 Tau at $\sim5.9\times 10^{34}$ erg and the flare with the lowest energy at $\sim1.3\times 10^{29}$ erg is seen in IL Aqr. The range of energies seen in the sample are comparable to energies seen in other work similar to this. Using Kepler data, \cite{hawley2014kepler} showed flares with energies in the range of $\sim2\times10^{28} - 2\times10^{33}$ erg on the M dwarf GJ 1243. Other stars in their sample were consistent with this but less extreme. In addition, \cite{lurie2015kepler} investigated the M5 binary system GJ 1245 also using 9 months of short cadence Kepler data, finding a total of 1288 flares on both stars with an energy range of $\sim1\times10^{30} - 1\times10^{33}$ erg.  For comparison, these large scale flares of order of $10^{34}$ erg are the equivalent to extremely large X-class flares, in terms of the solar flare GOES classification. In Table \ref{chi_squared_results} we note the median energy of the flares for each star.

\begin{table*}
\caption{Parallaxes along with their associated errors from {\sl Gaia} Data release 2 \citep{gai16,gai18} which are then inverted to calculate the distances to the stars in our sample. The errors on the quiescent luminosity includes the error on the distance and the PanStarrs magnitude. Three of our stars did not have parallaxes in the {\sl Gaia} DR2 catalogue and so these distances were taken from the EPIC Catalogue \citep{Huber2016} and marked with an asterisk.}
   \begin{center}
   \label{Gaia_data}

	\begin{tabular}{|c|c|c|c|c|}
    \hline 
    Name                       & Parallax           & Parallax Error & Distance         & $log(L_{star})$        \\ 
                               & mas                & mas            &  pc              & erg/s                  \\ 
    \hline
    201611969                  & 35.48              & 0.06           & 28.18 $\pm$ 0.04 & 31.679 $\pm$ 0.006     \\ 
    205204563                  & 68.78              & 0.12           & 14.54 $\pm$ 0.03 & 31.266 $\pm$ 0.007     \\ 
    205467732                  & --                 & --             & 28.3*            & 30.88                  \\ 
    206019387                  & 213.87             & 0.08           & 4.676$\pm$ 0.002 & 30.864 $\pm$ 0.001     \\ 
    206050032                  & 91.89              & 0.09           & 10.88 $\pm$ 0.01 & 30.564 $\pm$ 0.004     \\ 
    206053352                  & 25.3               & 0.2            & 39.47 $\pm$ 0.31 & 29.261 $\pm$ 0.032     \\
    206262336                  & 89.20              & 0.13           & 11.21 $\pm$ 0.02 & 30.385 $\pm$ 0.006     \\ 
    210317378                  & 28.59              & 0.06           & 34.97 $\pm$ 0.08 & 31.119 $\pm$ 0.009     \\
    210434433                  & 32.36              & 0.06           & 30.90 $\pm$ 0.06 & 31.188 $\pm$ 0.008     \\ 
    210460280                  & 27.61              & 0.04           & 36.22 $\pm$ 0.06 & 31.389 $\pm$ 0.006     \\ 
    210489654                  & 34.72              & 0.15           & 28.80 $\pm$ 0.12 & 31.011 $\pm$ 0.016     \\ 
    210579749                  & 58.01              & 0.05           & 17.24 $\pm$ 0.01 & 31.661 $\pm$ 0.003     \\ 
    210758829                  & 55.20              & 0.09           & 18.12 $\pm$ 0.03 & 30.330 $\pm$ 0.007     \\ 
    210764183                  & 34.12              & 0.42           & 29.31 $\pm$ 0.36 & 28.936 $\pm$ 0.049     \\ 
    210811310                  & 19.07              & 0.04           & 52.45 $\pm$ 0.12 & 31.477 $\pm$ 0.009     \\ 
    210894955                  & 34.69              & 0.13           & 28.83 $\pm$ 0.11 & 29.763 $\pm$ 0.015     \\ 
    211046195                  & 19.53              & 0.15           & 51.21 $\pm$ 0.40 & 29.988 $\pm$ 0.032     \\ 
    211069418                  & 8.13               & 0.65           & 122.9 $\pm$ 9.9  & 31.355 $\pm$ 0.321     \\ 
    211077349                  & 7.34               & 0.07           & 136.3 $\pm$ 1.3  & 31.306 $\pm$ 0.038     \\ 
    211082433                  & 12.62              & 0.17           & 79.24 $\pm$ 1.09 & 31.131 $\pm$ 0.055     \\ 
    211112686                  & 7.47               & 0.05           & 133.9 $\pm$ 0.9  & 31.623 $\pm$ 0.028     \\ 
    211117230                  & 7.58               & 0.05           & 131.9 $\pm$ 0.9  & 32.016 $\pm$ 0.027     \\ 
    211642294                  & 56.10              & 0.06           & 17.82 $\pm$ 0.02 & 31.307 $\pm$ 0.004     \\ 
    211945363                  & --                 & --             & 37.7*            & 31.41                  \\ 
    211970427                  & 5.28               & 0.15           & 189.3 $\pm$ 5.4  & 31.721 $\pm$ 0.113     \\ 
    212009427                  & 5.43               & 0.04           & 184.2 $\pm$ 1.2  & 32.083 $\pm$ 0.026     \\ 
    212029094                  & 5.39               & 0.06           & 185.3 $\pm$ 2.1  & 31.513 $\pm$ 0.045     \\ 
    212518629                  & 45.5               & 1.2            & 21.98 $\pm$ 0.58 & 30.832 $\pm$ 0.105     \\ 
    212776174                  & 41.31              & 0.05           & 24.21 $\pm$ 0.03 & 32.022 $\pm$ 0.005     \\ 
    212826600                  & --                 & --             & 109*             & 30.49                  \\
    228162462                  & 125.59             & 0.07           & 7.962$\pm$ 0.004 & 30.156 $\pm$ 0.002     \\ 
    \hline
    \end{tabular}

    \end{center}
\end{table*}

\section{Rotational Phase}
\label{rot_phase}

Stars with starspots can show a periodic change in brightness as the stars rotates due to the starspots being cooler than the surrounding photosphere. If stellar flares originate from the starspot, one might naively expect more flares would be seen at rotation minimum where the starspot is most visible. However, if a star has a low rotation angle (i.e. one of its rotation poles is close to being face on), spots near the pole would be visible at all phases, and flares would be seen at all rotation phases. To investigate this further, we now set out to determine the rotation phase of the flares we identified in the previous section.

For this analysis, stars with rotation periods shorter than the observation length were selected. Due to this, eight of the sources are omitted from this analysis and any further investigations due to the K2 lightcurve showing incomplete modulation which lead to the rotation period being unable to be confirmed. We phase folded and binned the lightcurves using the rotation period shown in Table \ref{Mdwarf_period} and the phase zero calculated previously, yielding rotation cycles covering all of the K2 lightcurve showing a minimum at $\phi$=0.0. We show the resulting phase folded and binned lightcurves in Figure \ref{lightphase}. Flares are present at practically all rotational phases for all stars. In many cases there are high energy flares present at rotation maximum, where we would expect the starspot to be least visible.  Many of the lightcurves show a roughly sinusoidal modulation suggesting the presence of one prominent starspot, although several sources, such as Wolf 1561 A (which is in a triple system) show evidence for a second starspot. The fact that all the stars shown in Figure \ref{lightphase} show a clear modulation suggests we are not observing them at low rotation angles.

In order to determine whether the phase distribution of the flares is random we used the $\chi^{2}$ statistic. Flares were split up into high and low energy with a cut-off determined by the median energy of all flares from each star. In addition, the rotational phase was split into 10 bins and $\chi^{2}$ was determined for each star in the low, high and all energy categories, where the degrees of freedom, $\nu$, is 9. Table \ref{chi_squared_results} shows the results for both high and low energy flares and also all flares in each star overall. Regarding low, high and all flare categories, none of the stars show a preference for rotational phase even at a 2$\sigma$ confidence level. Therefore, there is no evidence for the flares having any preference for rotational phase, which surprisingly indicates many flares may not originate from the large starspot. We now go on to investigate possible causes for this.

 \begin{table}
   \caption{For the stars shown in Figure \ref{lightphase} (with the exception of EPIC 210460280 which has only one flare) we show the $\chi^{2}_{\nu}$ value for whether each rotation phase bin (split into ten) had flares which were randomly distributed by phase. We split the flares into low and high energy where the cut off is determined by the median energy of all the flares for each star. None of the stars in our sample show a preference for flares at a certain rotational phase. }
   \begin{center}
     \begin{tabular}{ccccc}
      \hline 
     
     \multirow{2}{*}{EPIC} & median energy & \multicolumn{3}{c}{Reduced Chi-Squared} \\
        \cline{3-5}
         & (erg) & low & high & all \\
         \hline
         205204563 & 9.5 $\times 10^{30}$ & 0.78 & 2.00 & 1.00\\
         205467732 & 2.3 $\times 10^{31}$ & 1.17 & 1.17 & 0.55\\
         206262336 & 6.5 $\times 10^{30}$ & 0.92 & 1.13 & 0.61\\
         210317378 & 6.2 $\times 10^{31}$ & 0.78 & 1.22 & 0.97\\
         210434433 & 4.9 $\times 10^{31}$ & 0.89 & 0.89 & 1.22\\
         210579749 & 3.9 $\times 10^{31}$ & 0.89 & 0.89 & 0.67\\
         210758829 & 6.7 $\times 10^{30}$ & 1.14 & 0.30 & 0.62\\
         210764183 & 2.3 $\times 10^{31}$ & 1.44 & 1.00 & 1.11\\
         210811310 & 1.1 $\times 10^{31}$ & 1.00 & 1.00 & 2.00\\
         210894955 & 1.9 $\times 10^{31}$ & 1.10 & 1.35 & 0.47\\
         211046195 & 3.7 $\times 10^{32}$ & 0.33 & 0.78 & 0.78\\
         211069418 & 1.8 $\times 10^{32}$ & 0.93 & 1.33 & 0.65\\
         211077349 & 3.1 $\times 10^{32}$ & 0.82 & 0.82 & 1.15\\
         211082433 & 1.2 $\times 10^{32}$ & 1.11 & 0.39 & 0.72\\
         211112686 & 4.8 $\times 10^{32}$ & 0.58 & 1.21 & 1.18\\
         211117230 & 1.9 $\times 10^{33}$ & 0.87 & 0.99 & 0.69\\
         211642294 & 8.6 $\times 10^{31}$ & 0.78 & 0.89 & 1.00\\
         211970427 & 5.5 $\times 10^{32}$ & 0.82 & 0.82 & 1.24\\
         212009427 & 7.8 $\times 10^{32}$ & 0.54 & 0.81 & 0.33\\
         212029094 & 2.5 $\times 10^{31}$ & 1.00 & 0.89 & 1.52\\
         212776174 & 4.3 $\times 10^{31}$ & 1.22 & 0.78 & 0.97\\
         228162462 & 4.6 $\times 10^{30}$ & 0.83 & 0.48 & 0.94\\
         \hline
      \end{tabular}    
    \end{center}
    \label{chi_squared_results}
\end{table}

\section{Discussion}

We have analysed flaring activity from a sample of 31 M dwarfs covering a range of spectral types using K2 short cadence data. We derived (or placed lower limits on) the rotation periods of 29 of these stars using the K2 lightcurves, many for the first time. In addition to this we used the flare characteristics (energy, duration and phase of the rotation cycle) to compute a statistical analysis of flares of this sample.

It is known, from previous studies \citep[e.g.][]{mohanty2003rotation,mclean2012radio}, that faster rotating stars show greater flaring activity. Moreover, activity drops for stars with rotation periods $>$ 10 days \citep{stelzer2016rotation}. Our findings are  consistent with these results. However, our work has identified an area which has not previously been studied in great detail. If flares originate from the same starspot which causes the rotational modulation, we would expect to observe a clear correlation of flares with rotational phases: this is not what we observe. None of the stars in our sample show evidence for flares being preferentially seen at certain rotation phases. This result is unexpected and seems to point to the conclusion that the majority of flares do not originate from the prominent starspot.

Where do flares originate on these active stars and how are they generated? We consider three possible scenarios. Firstly, there is the potential of magnetic interaction with a second star in a binary system. It is possible for interactions between the M dwarf and a binary companion causing increased magnetic activity between the stars and in turn the generation of flares at locations other than a dominant starspot. As summarised by \cite{kouwenhoven2009exploring}, 30--40 \% of M dwarfs are members of a binary system. For late M dwarfs and brown dwarfs this drops to 10 - 30 \%. This suggests less than a dozen of our sample will be in a binary system and hence binarity is unlikely to play a prominent role in resolving this question. 

Our second scenario is that magnetic interaction could occur with a planet orbiting the M dwarf. \cite{dressing2015occurrence} present an updated occurrence rate for planets orbiting early M dwarfs as 2.5 planets per M dwarf star. Depending on the number of planets orbiting the host star (and the mass, radius and magnetic field of the planet), it could be induced magnetic activity between the star-planet system which causes the increased flaring activity. However, both the first and second scenario would depend on a small separation between the orbiting star or planet to allow for any magnetic interaction.

We now consider the likely separation between the photosphere of the M dwarf and the magnetic interaction region. In the multi-scale field scenario of \cite{yadav2015formation}, only the large scale field (lower order multiple) would interact with any orbiting planets and the location of any `null points' (where the global topology of the magnetic field changes) would be in a transitional region where the long range dipole field becomes weaker than the quadrupole (higher order) field. As the null point induced by the planet is also associated with the dipole field it cannot be too close to the stellar surface. Moreover, this kind of null point would not be associated with the dominant stellar spot (which is not created by the dipole field component). Furthermore, the reconnection region would not easily be eclipsed by the star so, it is natural that we do not see any correlation with the phases and the flares.

A third possibility is the presence of polar spots on the M dwarf. Depending on the viewing geometry and the relative inclination of the rotational and magnetic moment axes of the star, polar spots could be seen at all phases, interacting with emerging active regions and spot free regions as the star rotates, causing continuously visible flaring activity. The Sun does not posses polar spots, so the presence of polar spots on M dwarfs would support the view that the generation of the magnetic field in these stars differs from than of the Sun.

Despite the absence of polar spots on the Sun, \cite{schrijver2001formation} model the formation of polar spots on rapidly rotating (6 d) Sun-like stars due to the poleward migration of the magnetic field. Their models predicted very active stars possessing polar caps with topologies of one polarity encircling another. This magnetic configuration could generate very large filaments, flares and Coronal Mass Ejections. Despite the differences between rapidly rotating Sun-like stars and fully convective low mass stars, the work of \cite{schrijver2001formation} highlighted that a global poloidal field and a shear underneath/next to it in a perpendicular direction, a local dynamo can occur. Hence, this could lead to flare activity.

We now go on to consider the formation of polar spots on fully convective low mass stars. \cite{yadav2015formation} investigated the conditions necessary for formation of polar spots in convection driven dynamos. This directly applies to many stars in our sample as they are fully convective, and a magnetic field driven by the $\alpha^{2}$ dynamo mechanism. As a result of their parameter study, they determine three key features for large spot formation in fully convective stars: i) rotation driven convection, ii) many scale heights in the convection zone, iii) a dynamo producing an axial-dipole field. All of our stars posses these properties and so, could this be the solution to the key problem noticed in the rotational phase of the flares? 

\cite{yadav2015formation} also show a self-consistent distributed dynamo can spontaneously generate high-latitude dark spots when a large-scale magnetic field, generated in the bulk of the convection zone, interacts with and locally quenches flow near the surface. This is similar to findings reported by \cite{schrijver2001formation}, who explored the migration of surface magnetic fields towards the poles. Rapid rotation is vital for the formation of such dark polar spots. However, if there is a global polaroidal field and a shear underneath/next to it in a perpendicular direction, a local dynamo can occur.

Such flare activity from the polar regions could be caused by large-scale 2D vortices within the poloidal fields which thread through the temperature inversion layer (in the polar regions) acting in a similar way to charged particles when they experience a force across the field lines perpendicular to their motion. When such 2D vortices are formed they may wind up the field line with them and when these eddies encounter the rim of the polar cap, (whose axis is not aligned with the rotation axis of the star), magnetic reconnection may occur generating flares. 

We have outlined three scenarios to explain the lack of rotational phase preference for flares in our M dwarf sample. In order to test our theories further we would need to compare results from stars with low and high inclination. \cite{Davenport2015} and \cite{silverberg2016} reported observations of GJ 1243 which has a high rotation inclination and a high latitude spot. 
With many thousands of flares being detected, no correlation with its 0.59 day rotational period was found. A sample of stars with well defined inclination values is the next step in allowing us to distinguish between the three scenarios to explain our main result.

\section{Conclusions}

Previous observations of activity levels in fast and slow rotators suggest a rotation-dependent transition in the magnetic properties of the atmosphere of M dwarfs, where the transition corresponds to approximately 10 days. Using K2 SC observations of a sample of 34 M dwarfs, we have found an interesting result. There is no correlation between the rotation phase and the number of flares. Given these stars all show significant rotational modulation amplitude due to a starspot, this is a surprise. 

New wide field surveys which are red sensitive will be suited to exploring these issues in greater detail and with larger sample sizes. For instance, the New Generation Transit Survey (NGTS) \citep{wheatley2017next} has a field of view of 96 square degree and is red sensitive. Although its prime goal is the detection of Neptune and super-Earth size exo-planets it will obtain long duration lightcurves of many red dwarfs. The Transiting Exoplanet Survey Satellite  (TESS) \citep{Ricker2015} was launched on the 18th April 2018. TESS will be sensitive to stars brighter than $V\sim$12 and will have 27 day observation blocks covering 24$^{\circ}\times96^{\circ}$ of sky with a cadence of 1 min for many objects. It will therefore be suitable to address our key questions. Furthermore, TESS will provide photometric precision which is one order of magnitude greater than Kepler/K2, therefore, allowing a better evaluation of the low-energy part of the flare frequency/flare energy distribution. 

\section*{Acknowledgements}
We thank Andrew Vanderberg for kindly detrended these K2 short cadence data and the K2 Guest Observer Office Team for their support and enthusiasm for the mission. Armagh Observatory and Planetarium is core funded by the Northern Ireland Government through the Dept. for Communities. LD acknowledges funding from an STFC  studentship. 
This work presents results from the European Space Agency (ESA) space mission {\sl Gaia}. {\sl Gaia} data is being processed by the {\sl Gaia} Data Processing and Analysis Consortium (DPAC). Funding for the DPAC is provided by national institutions, in particular the institutions participating in the {\sl Gaia} MultiLateral Agreement (MLA). The Gaia mission website is \url{https://www.cosmos.esa.int/gaia}. The Gaia archive website is \url{https://archives.esac.esa.int/gaia}.

The Pan-STARRS1 Surveys (PS1) and the PS1 public science archive have been made possible through contributions by the Institute for Astronomy, the University of Hawaii, the Pan-STARRS Project Office, the Max-Planck Society and its participating institutes, the Max Planck Institute for Astronomy, Heidelberg and the Max Planck Institute for Extraterrestrial Physics, Garching, The Johns Hopkins University, Durham University, the University of Edinburgh, the Queen's University Belfast, the Harvard-Smithsonian Center for Astrophysics, the Las Cumbres Observatory Global Telescope Network Incorporated, the National Central University of Taiwan, the Space Telescope Science Institute, the National Aeronautics and Space Administration under Grant No. NNX08AR22G issued through the Planetary Science Division of the NASA Science Mission Directorate, the National Science Foundation Grant No. AST-1238877, the University of Maryland, Eotvos Lorand University (ELTE), the Los Alamos National Laboratory, and the Gordon and Betty Moore Foundation. 

\bibliographystyle{mnras}
\bibliography{K2_paper.bib} 

\appendix

\section{PanStarrs Magnitudes}
\begin{table*}
\caption{For all the stars in our sample we present the corresponding PanStarrs magnitudes \citep{chambers2016panstarrs} in the g, r i, and z bands with errors which are also taken from the PanStarrs catalogue. These magnitudes are used to create a template spectrum of each star and used in the calculation of the quiescent Kepler luminosity.}

   \begin{center}
   \label{Panstarrs_mag}

	\begin{tabular}{|c|c|c|c|c|c|c|c|}
    \hline 
	Name                  & EPIC ID     &  g                        & r                       & i                       & z                       \\
    \hline
	LHS 2420              & 201611969   &  $13.0682 \pm 0.0010$     & $11.7446 \pm 0.0010$    & $10.0773$               & $10.6904 \pm 0.1852$    \\
	LP 804-27             & 205204563   &  $12.2986 \pm 0.0010$     & $11.0668 \pm 0.1391$    & $9.8560 \pm 0.1723$     & $9.0415 \pm 0.0378$     \\
	GJ 3954               & 205467732   &  $15.0310 \pm 0.0034$     & $13.7505 \pm 0.0029$    & $12.2592 \pm 0.3998$    & $11.1509 \pm 0.0210$    \\
	IL Aqr                & 206019387   &  $11.0744 \pm 0.0150$     & $8.9746 \pm 0.0692$     & $8.4762 \pm 0.0394$     &  $8.1422 \pm 0.2303$    \\
	LP 760-3              & 206050032   &  $12.2150$                & $11.5380$               & $11.2720$               & $11.1370$               \\
    2MASSI J2214-1319     & 206053352   &  $20.8604 \pm 0.0406$     & $19.4495 \pm 0.0081$    & $16.9408 \pm 0.0041$    & $15.7102 \pm 0.0042$    \\
	Wolf 1561 A           & 206262336   &  $14.2094 \pm 0.0001$     & $13.5603 \pm 0.0784$    & $11.3962 \pm 0.0352$    & $10.4803 \pm 0.0635$    \\
	HG 7-26               & 210317378   &  $14.3808 \pm 0.0065$     & $13.2037 \pm 0.0107$    & $12.1530 \pm 0.0701$    & $11.4553 \pm 0.0371$    \\
	NLTT 12593            & 210434433   &  $13.6522 \pm 0.0005$     & $12.3198 \pm 0.0010$    & 11.9320                 & $10.7620 \pm 0.0397$    \\
	G 6-33                & 210460280   &  $12.5168 \pm 0.0055$     & $11.6516 \pm 0.0404$    & $12.2169 \pm 0.0385$    & $11.8217 \pm 0.0991$    \\
	LP 415-363            & 210489654   &  $14.4373 \pm 0.0057$     & $13.3156 \pm 0.0037$    & $11.8085 \pm 0.0385$    & $13.3588 \pm 0.0012$    \\
	MCC 428               & 210579749   &  $11.2964 \pm 0.0520$     & $10.0182 \pm 0.0124$    & $9.3745 \pm 0.2068$     & $8.5127 \pm 0.0229$     \\
	GJ 3225               & 210758829   &  $15.5532 \pm 0.0024$     & $14.2847 \pm 0.0037$    & $12.5702 \pm 0.0183$    & $11.9246 \pm 0.1065$    \\
	2MASS J0326+1919      & 210764183   &  $21.4785 \pm 0.1072$     & $19.7707 \pm 0.0250$    & $17.1323 \pm 0.0032$    & $15.7071 \pm 0.0027$    \\
	LP 414-108            & 210811310   &  $13.7311 \pm 0.0100$     & $12.5716 \pm 0.0599$    & $12.3544 \pm 0.1876$    & $12.2010 \pm 0.3246$    \\
	LP 357-206            & 210894955   &  $18.4372 \pm 0.0077$     & $17.1264 \pm 0.0034$    & $14.9935 \pm 0.0013$    & $14.0026 \pm 0.0041$    \\
	2MASS J0335+2342      & 211046195   &  $19.5600 \pm 0.0100$     & $18.2366 \pm 0.0054$    & $15.6818 \pm 0.0029$    & $14.4760 \pm 0.0047$    \\
	LT Tau                & 211069418   &  $16.8460 \pm 0.0065$     & $15.6460 \pm 0.0035$    & $14.2525 \pm 0.0012$    & $13.3161 \pm 0.0017$    \\
	V497 Tau              & 211077349   &  $17.0385 \pm 0.0053$     & $15.7580 \pm 0.0126$    & $14.5928 \pm 0.0049$    & $14.0379 \pm 0.0036$    \\
	V692 Tau              & 211082433   &  $16.7380 \pm 0.0118$     & $15.5423 \pm 0.0052$    & $13.8039 \pm 0.0004$    & 12.8970                 \\
	V631 Tau              & 211112686   &  $15.9072 \pm 0.0060$     & $14.7119 \pm  0.0089$   & $13.8246 \pm 0.0100$    & $13.3578 \pm 0.0050$    \\
    V* MY Tau             & 211117230   &  $14.6712 \pm 0.0068$     & $13.5257 \pm 0.0010$    & $12.8800$               & $12.4340$               \\
	GJ 3508               & 211642294   &  $12.3840 \pm 0.0010$     & $11.7487 \pm 0.0461$    & $10.0195 \pm 0.0502$    & $11.4547 \pm 0.0000$    \\
	LP 426-35             & 211945363   &  $14.0742 \pm 0.0024$     & $13.3732 \pm 0.4861$    & $11.3541 \pm 0.0010$    & $11.9661 \pm 0.0010$    \\
	AX Cnc                & 211970427   & $16.6625 \pm 0.0041$      & $15.4590 \pm 0.0008$    & $14.2677 \pm 0.0010$    & $13.7382 \pm 0.0034$    \\
	2MASS J0831+2024      & 212009427   &  $15.3423 \pm 0.0041$     & $14.1807 \pm 0.0014$    & 13.3870                 & $13.0250 \pm 0.0079$    \\
	2MASS J0839+2044      & 212029094   & $17.0009 \pm 0.0051$      & $15.7975 \pm 0.0017$    & $14.7718 \pm 0.0023$    & $14.3082 \pm 0.0030$    \\
	LP 737-14             & 212518629   &  $14.3029 \pm 0.0051$     & $13.0944 \pm 0.0010$    & $11.7148 \pm 0.0400$    & $12.1497 \pm 0.0010$    \\
	BD-05 3740            & 212776174   &  $11.4172 \pm 0.0406$     & $9.2114 \pm 0.0875$     & $9.3707 \pm 0.0523$     & $9.1667 \pm 0.0487$     \\
    2MASSI J1332-0441     & 212826600   &  $20.2436 \pm 0.0221$     & $18.7220 \pm 0.0132$    & $16.0798 \pm 0.0023$    & $14.7831 \pm 0.0025$    \\
	GJ 1224               & 228162462   &  $14.1926 \pm 0.0010$     & $13.0493 \pm 0.0098$    & $11.1694 \pm 0.0134$    & 10.8942                 \\
    \hline
    \end{tabular}

    \end{center}
\end{table*}

\newpage
\section{Lightcurves}

\begin{figure*}
\centering
\subfloat[]{\includegraphics[width=.35\textwidth]{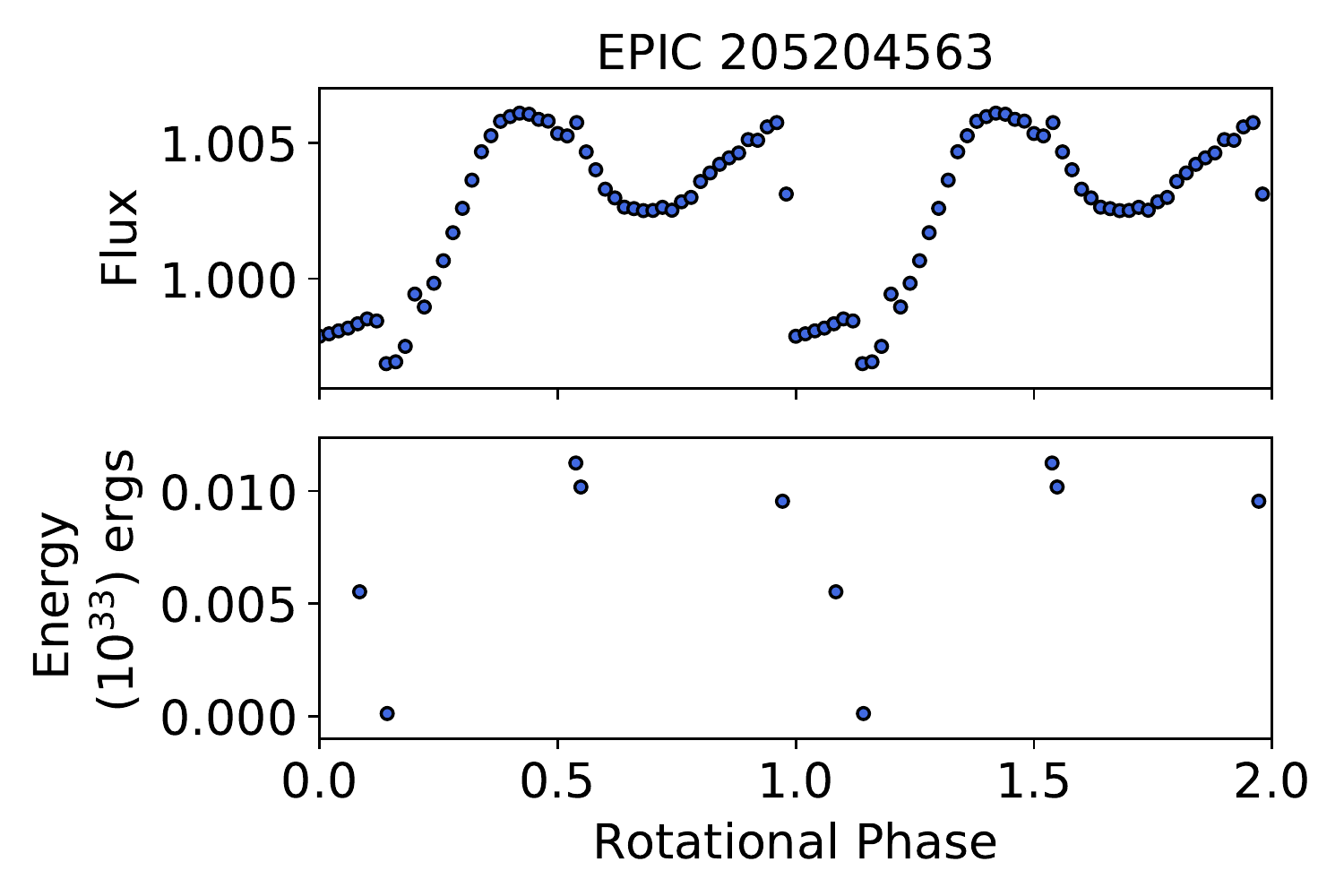}}
\subfloat[]{\includegraphics[width=.35\textwidth]{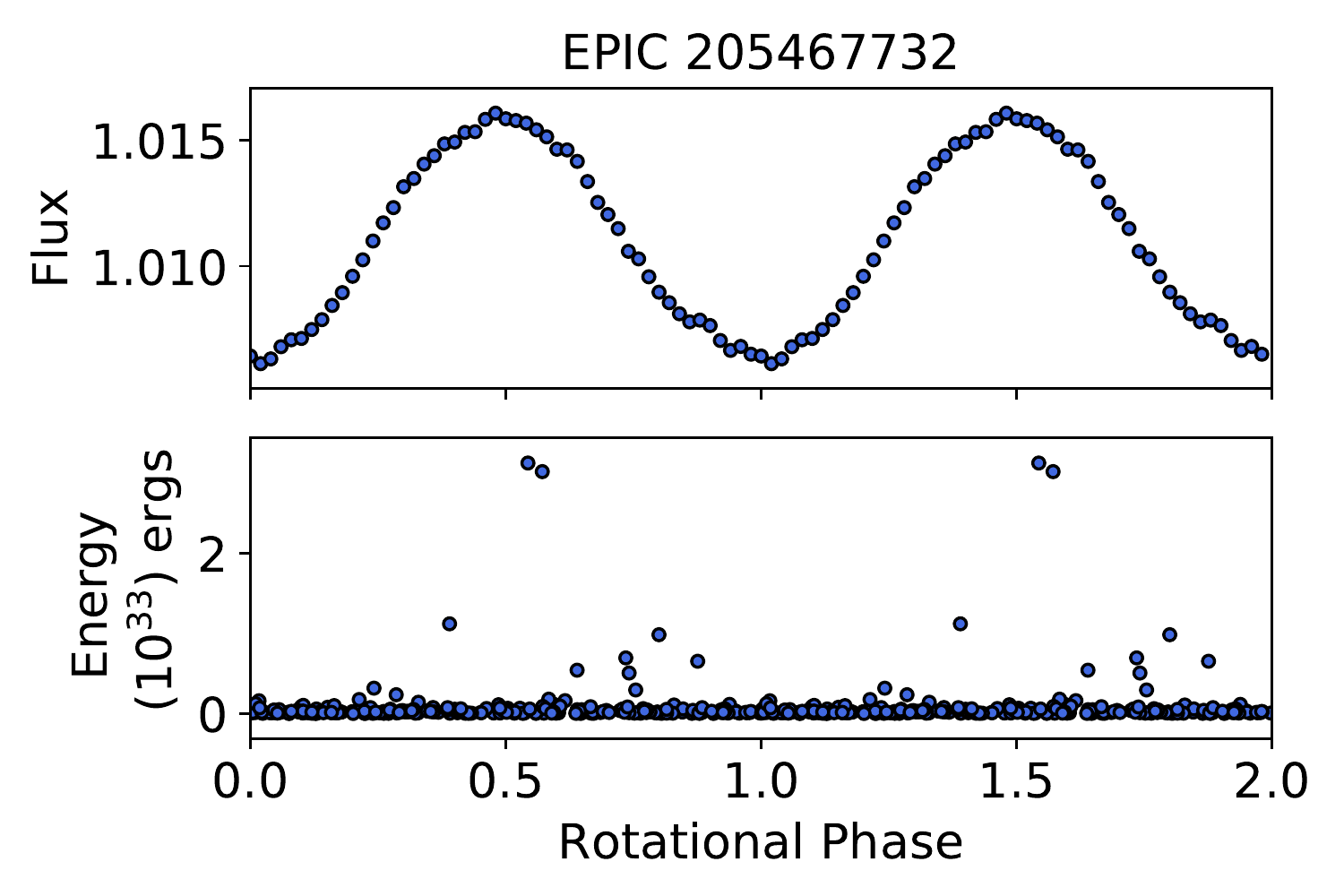}}
\subfloat[]{\includegraphics[width=.35\textwidth]{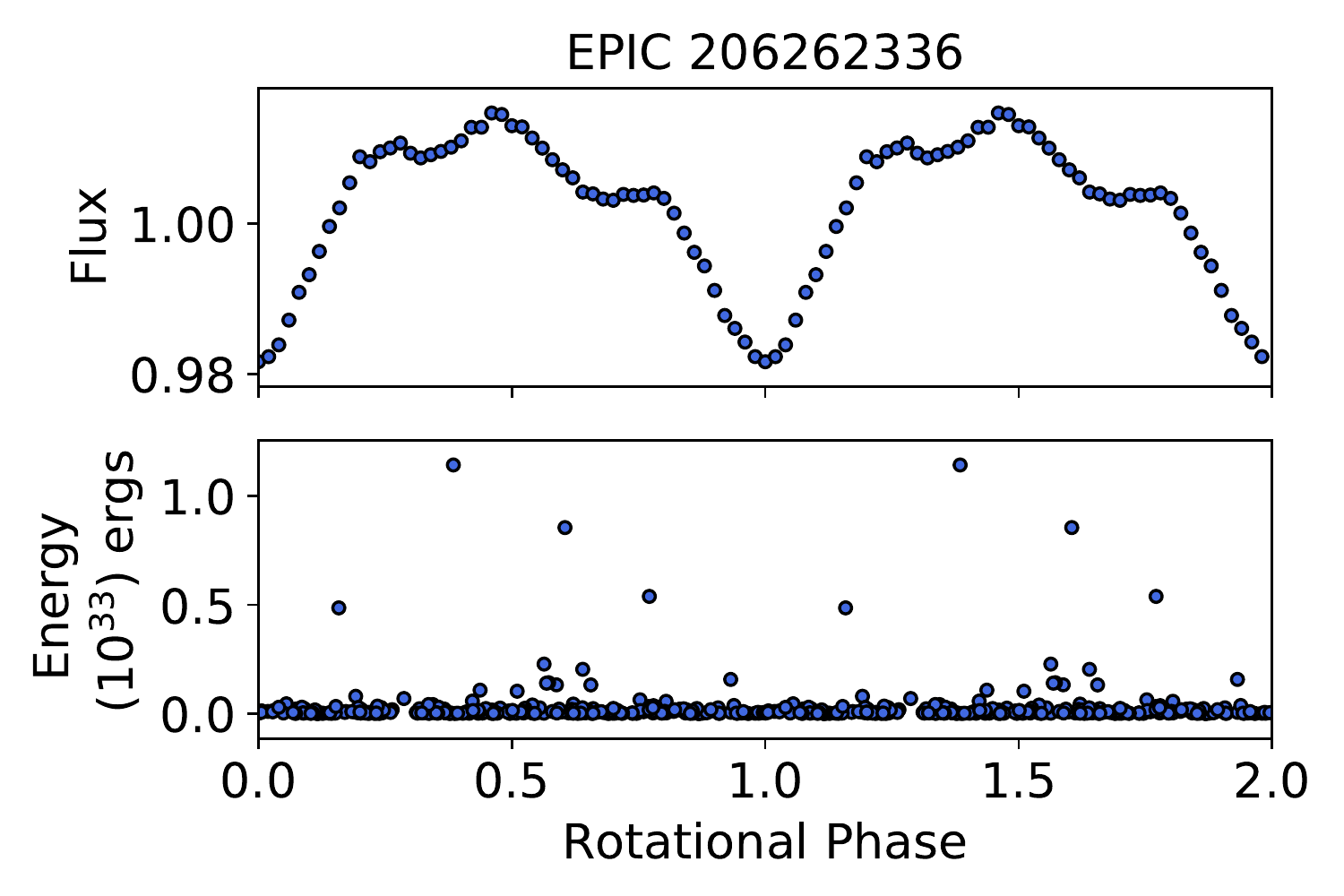}}\\
\subfloat[]{\includegraphics[width=.35\textwidth]{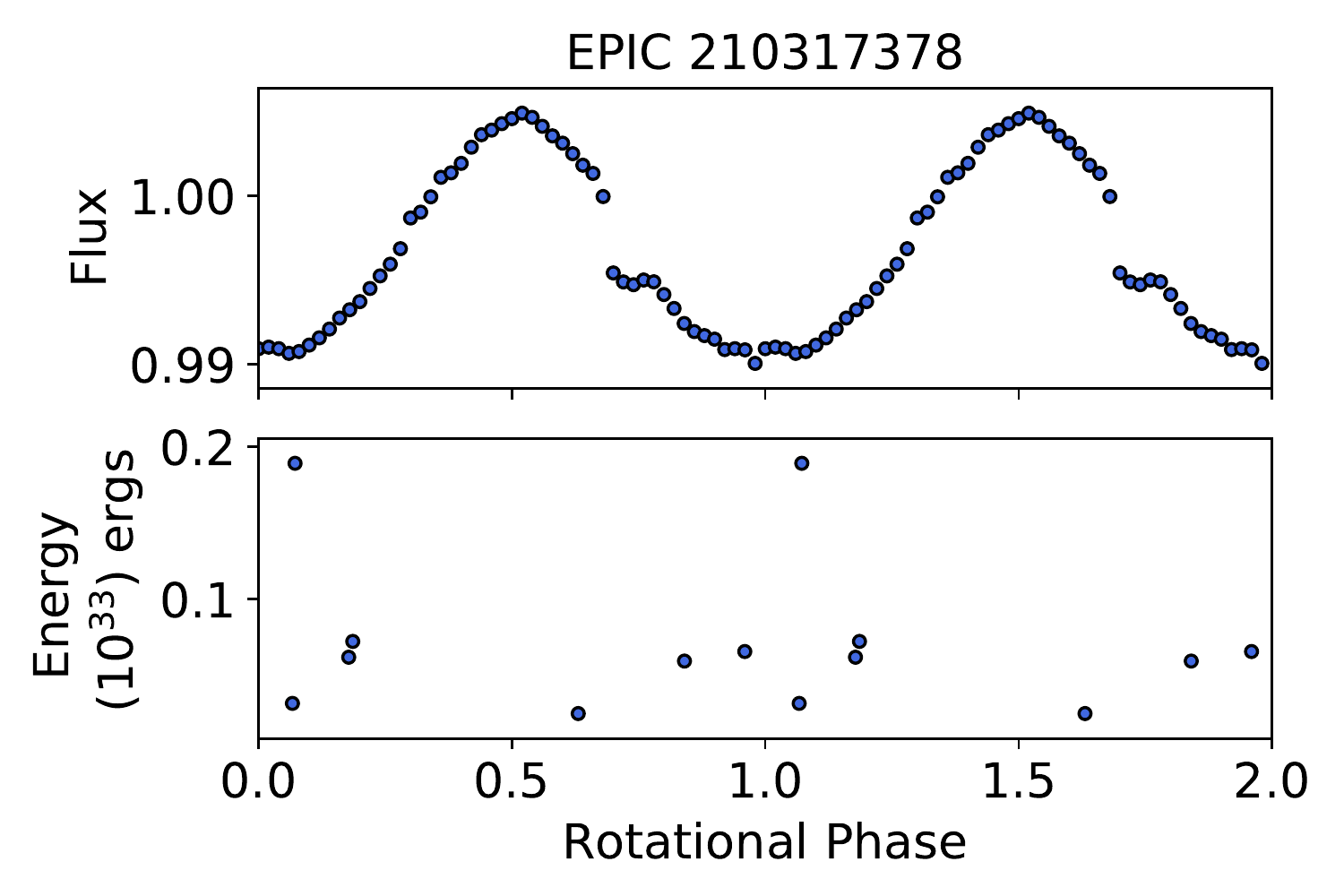}}
\subfloat[]{\includegraphics[width=.35\textwidth]{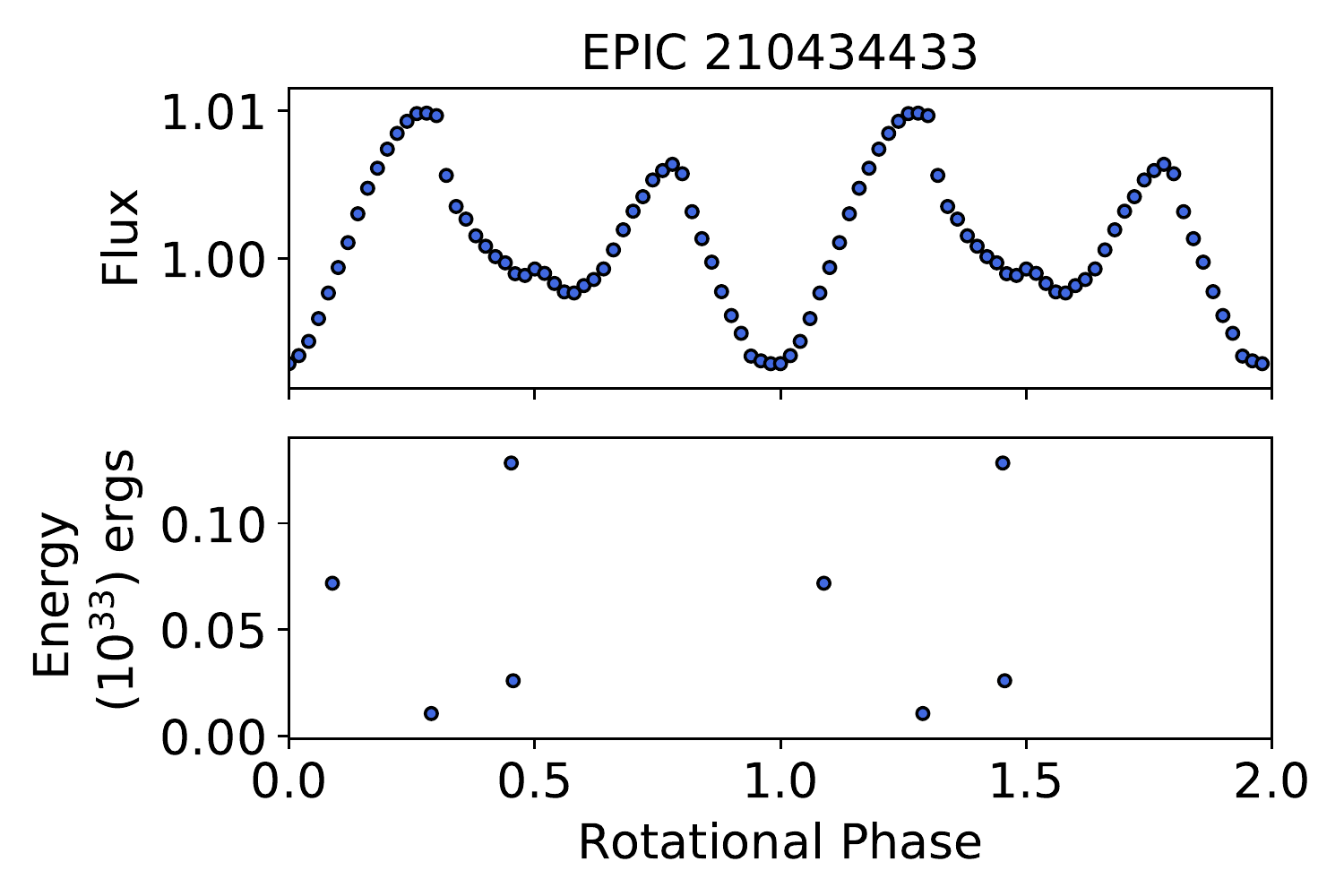}}
\subfloat[]{\includegraphics[width=.35\textwidth]{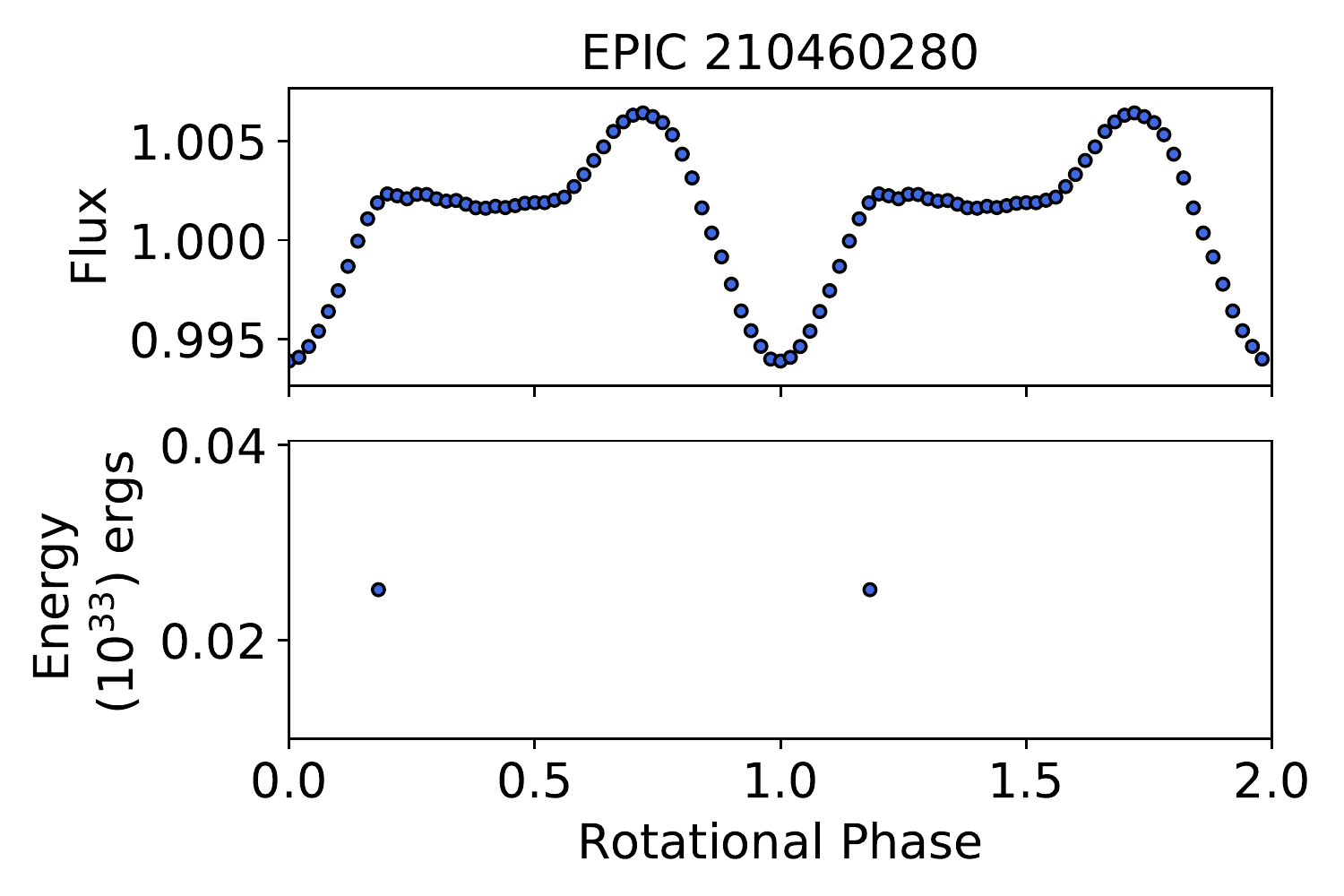}}\\
\subfloat[]{\includegraphics[width=.35\textwidth]{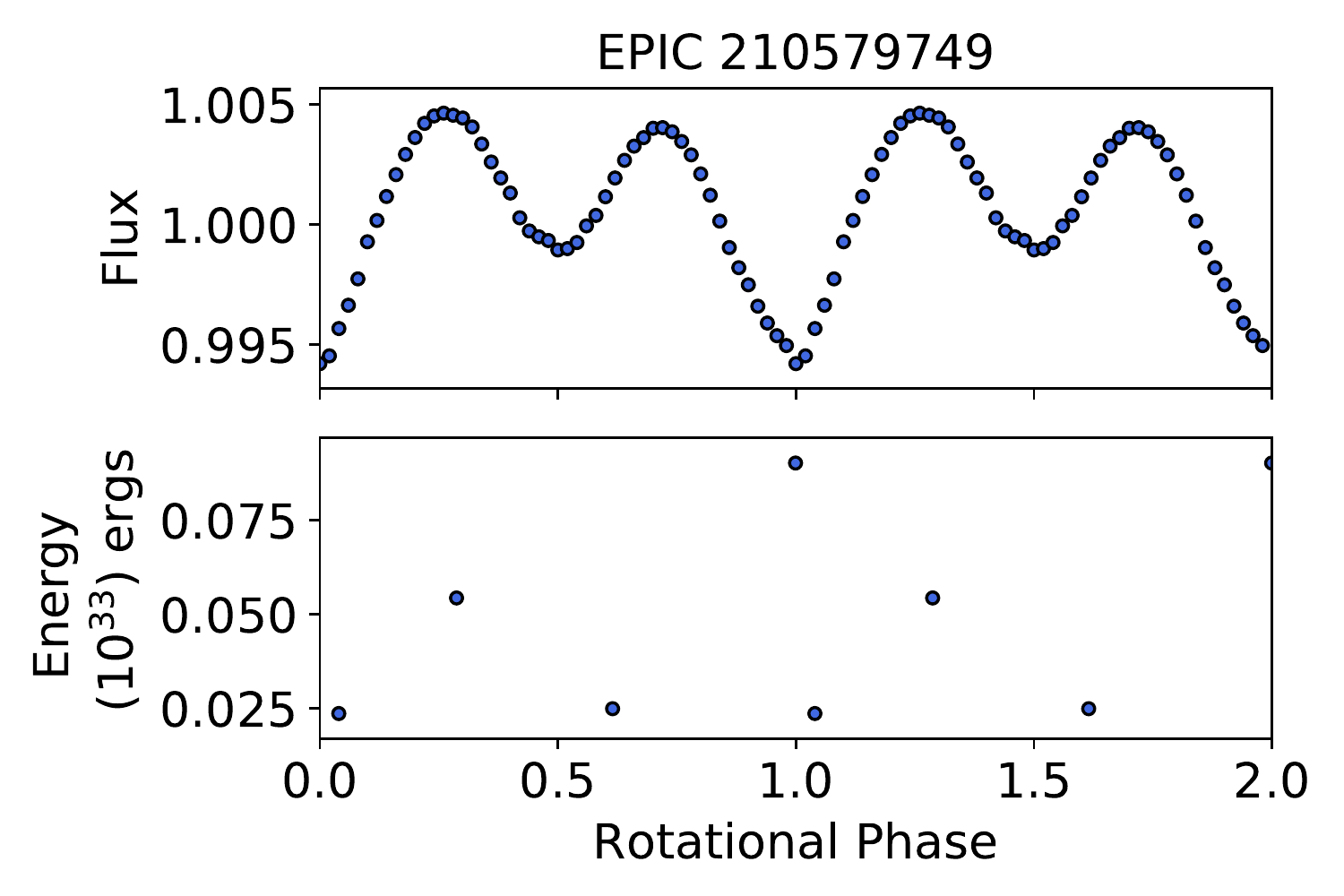}}
\subfloat[]{\includegraphics[width=.35\textwidth]{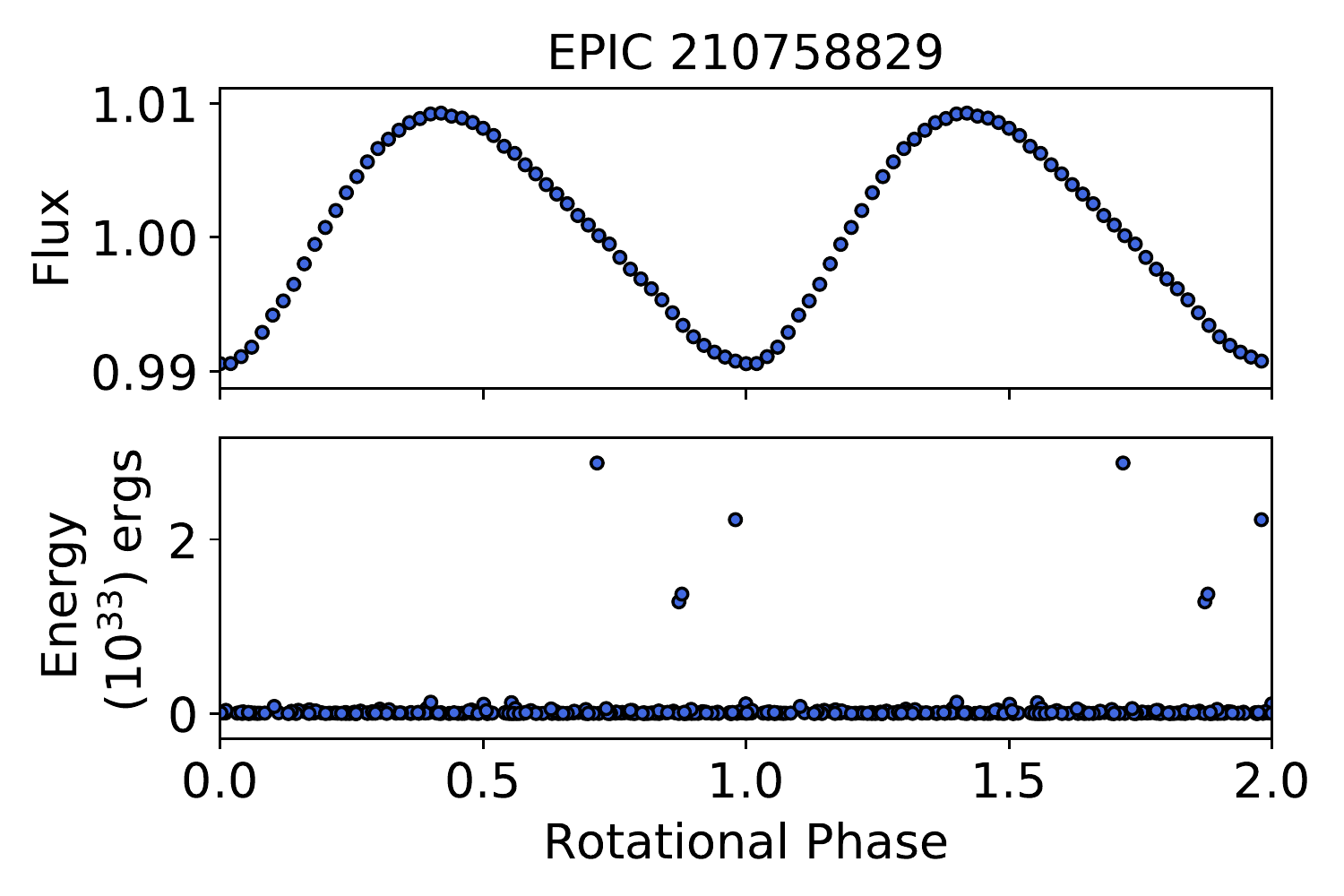}}
\subfloat[]{\includegraphics[width=.35\textwidth]{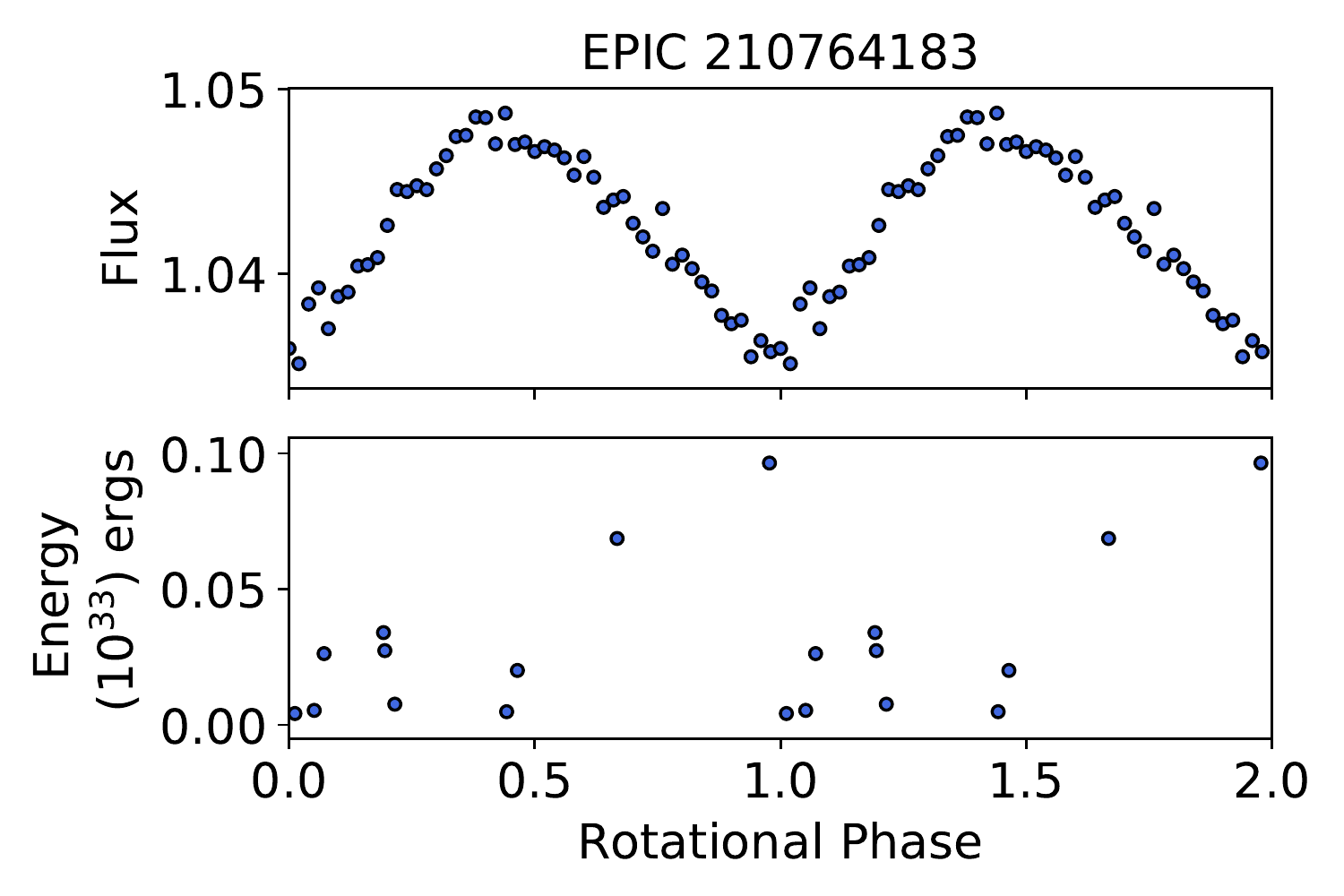}}\\
\subfloat[]{\includegraphics[width=.35\textwidth]{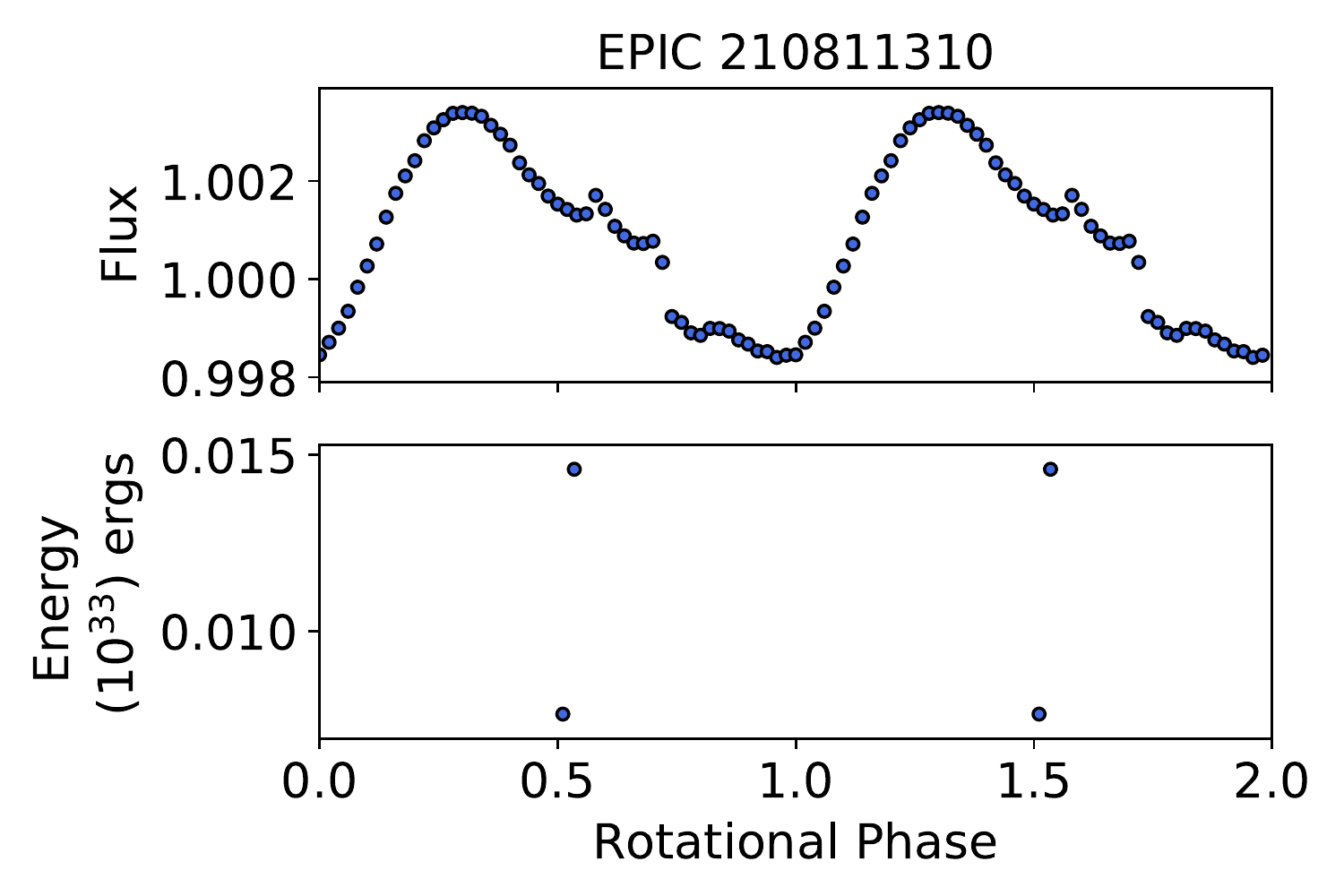}}
\subfloat[]{\includegraphics[width=.35\textwidth]{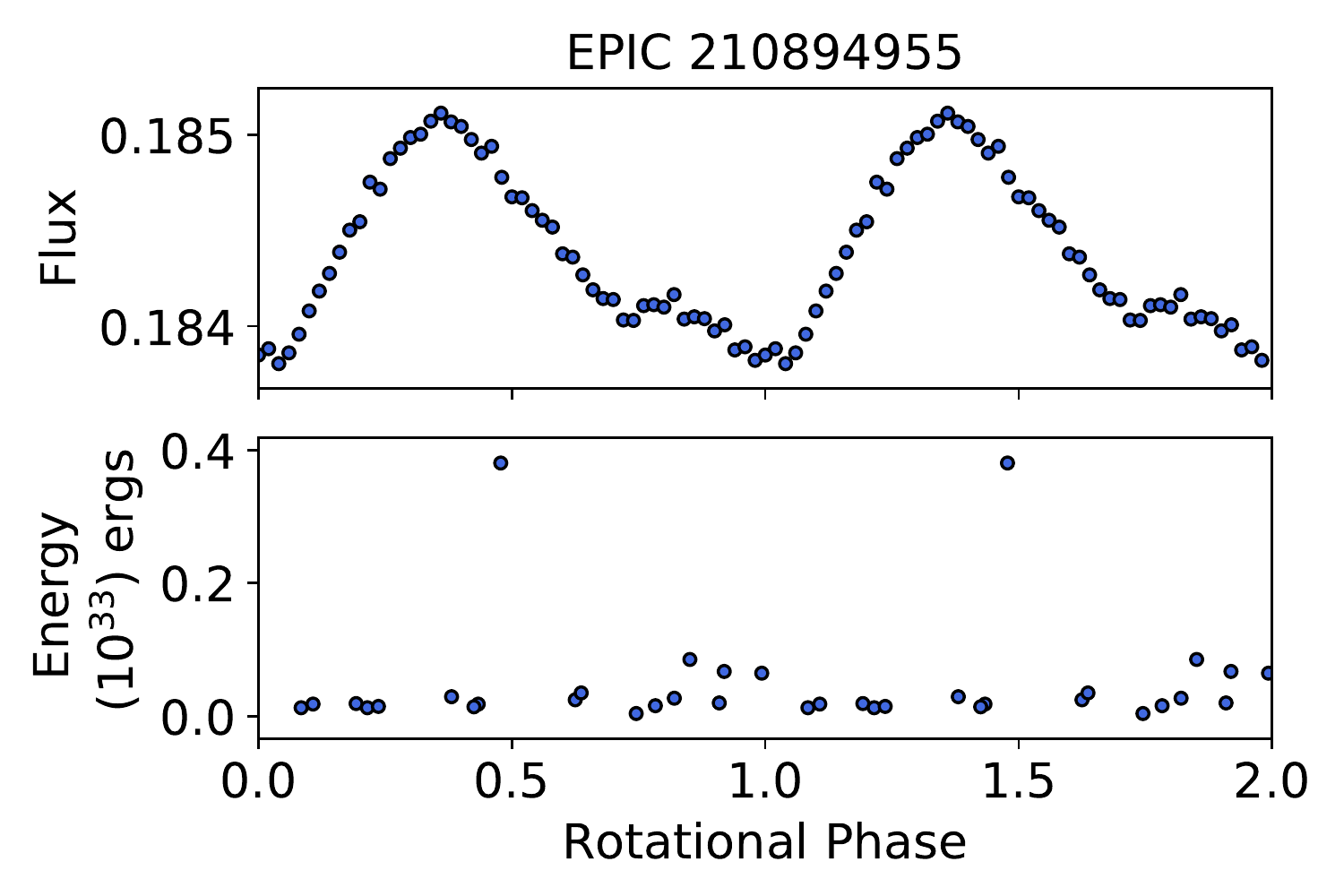}}
\subfloat[]{\includegraphics[width=.35\textwidth]{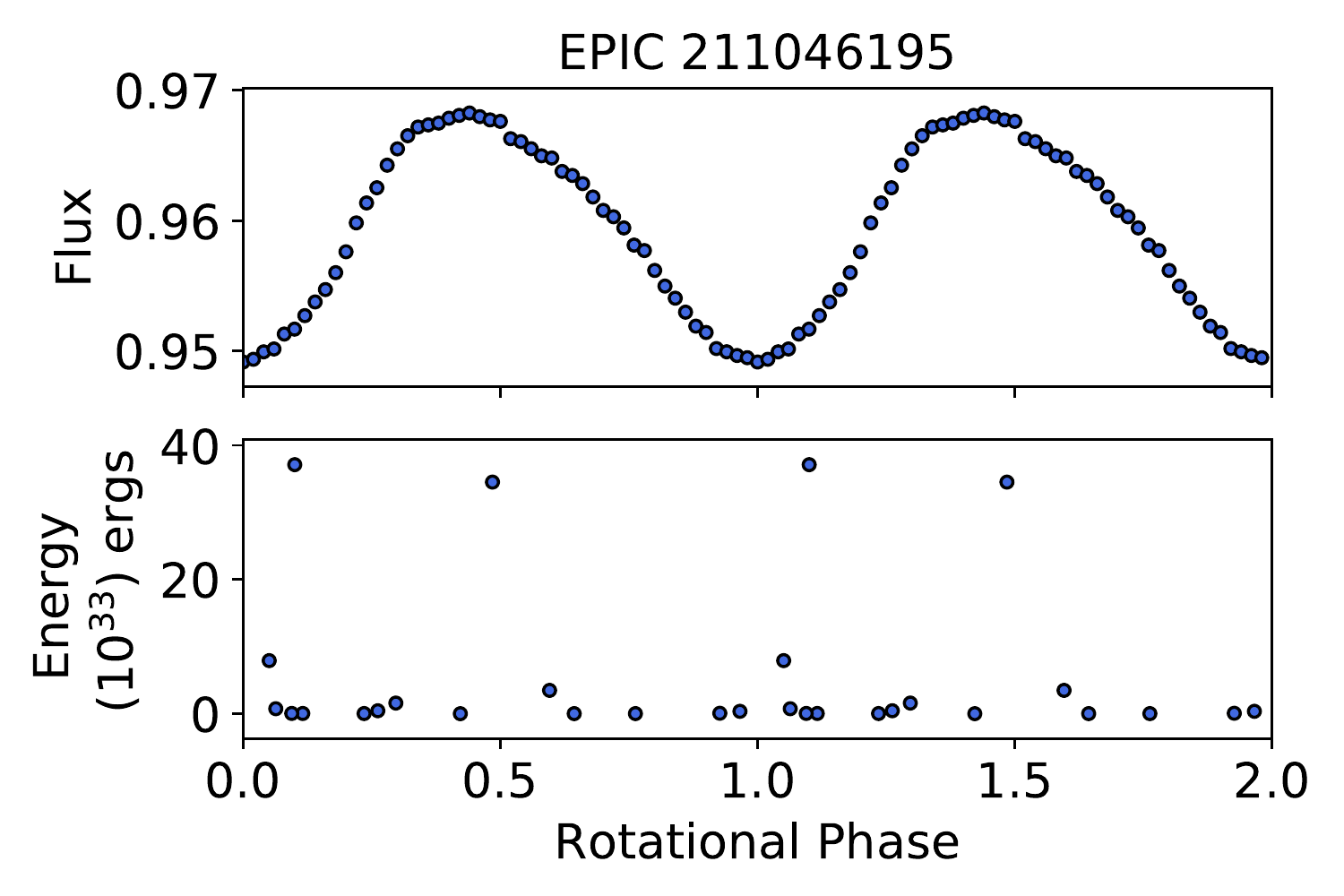}}\\
\caption{In the top panels we show the lightcurves of each star phased and binned on the rotation period such that there are 50 bins per rotation phase. In the bottom panels we show the phase of the flares with the energy. For each star we plot the data twice so they cover rotation phase 0.0--2.0 where 1.0--2.0 is simply a repeat of 0.0--1.0. The star shown in panel (a) shows three rotation cycles with a long term trend and the variation may indicate significant cycle to cycle variability.}
\label{lightphase}
\end{figure*}

\begin{figure*}\ContinuedFloat
\centering
\subfloat[]{\includegraphics[width=.35\textwidth]{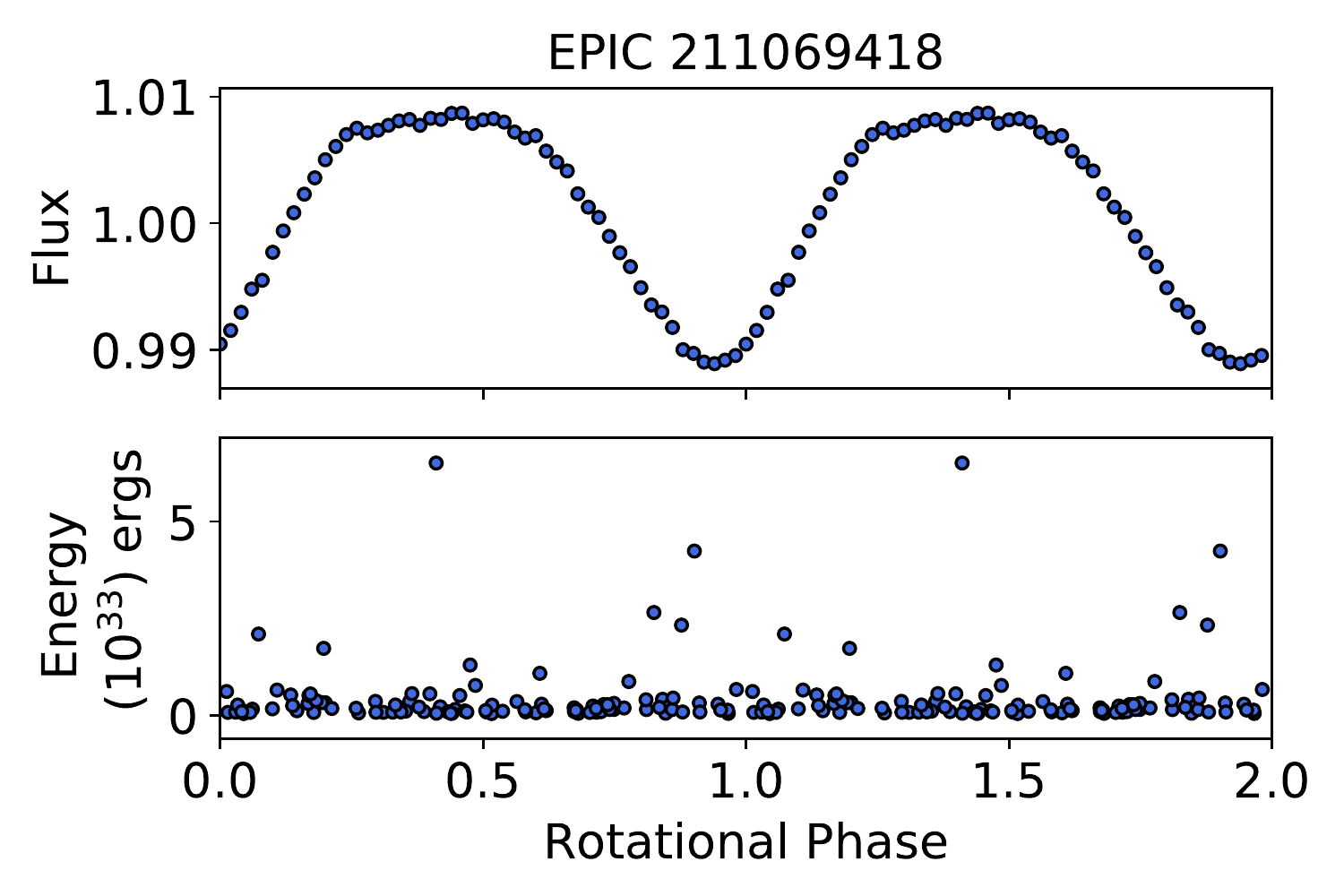}}
\subfloat[]{\includegraphics[width=.35\textwidth]{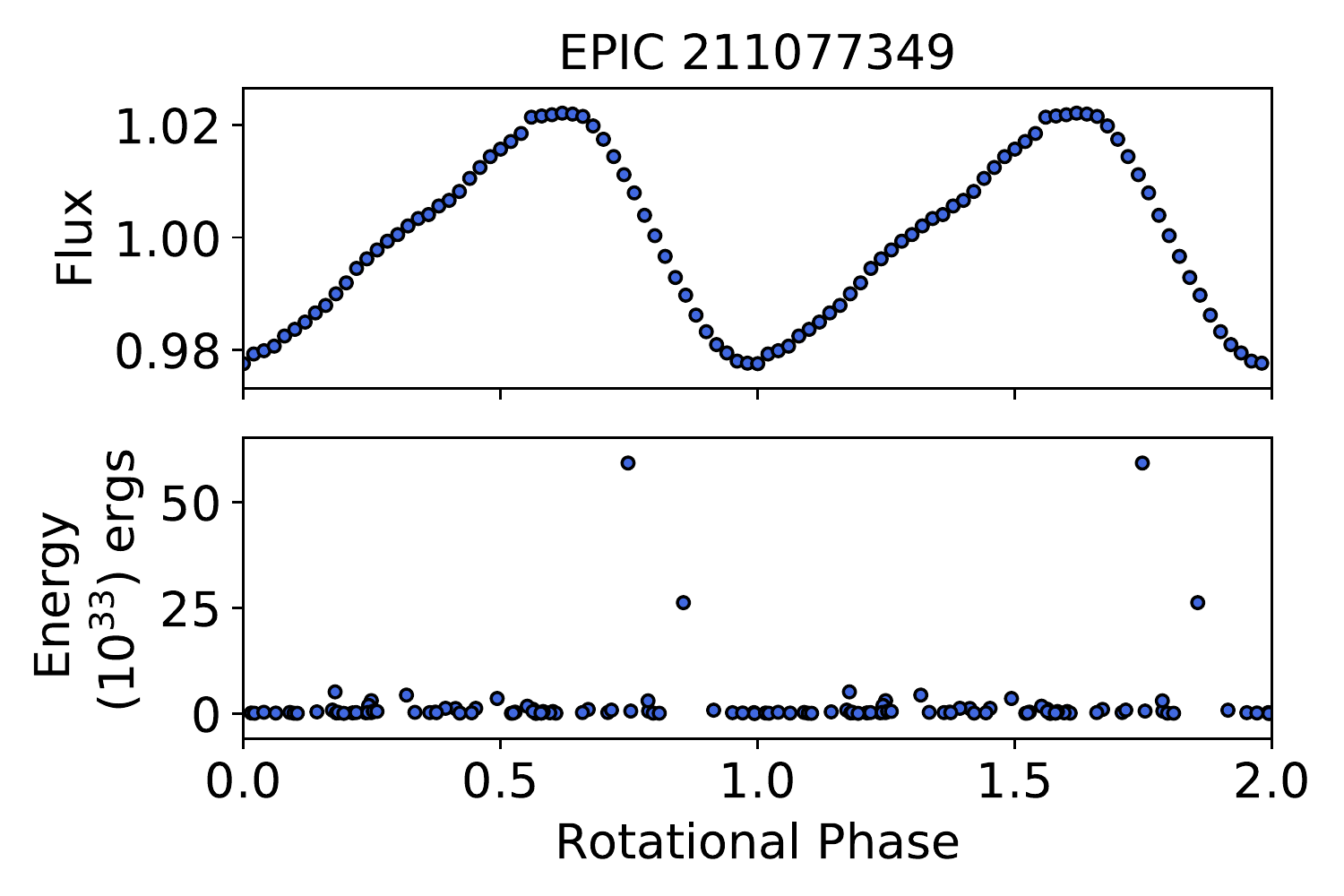}}
\subfloat[]{\includegraphics[width=.35\textwidth]{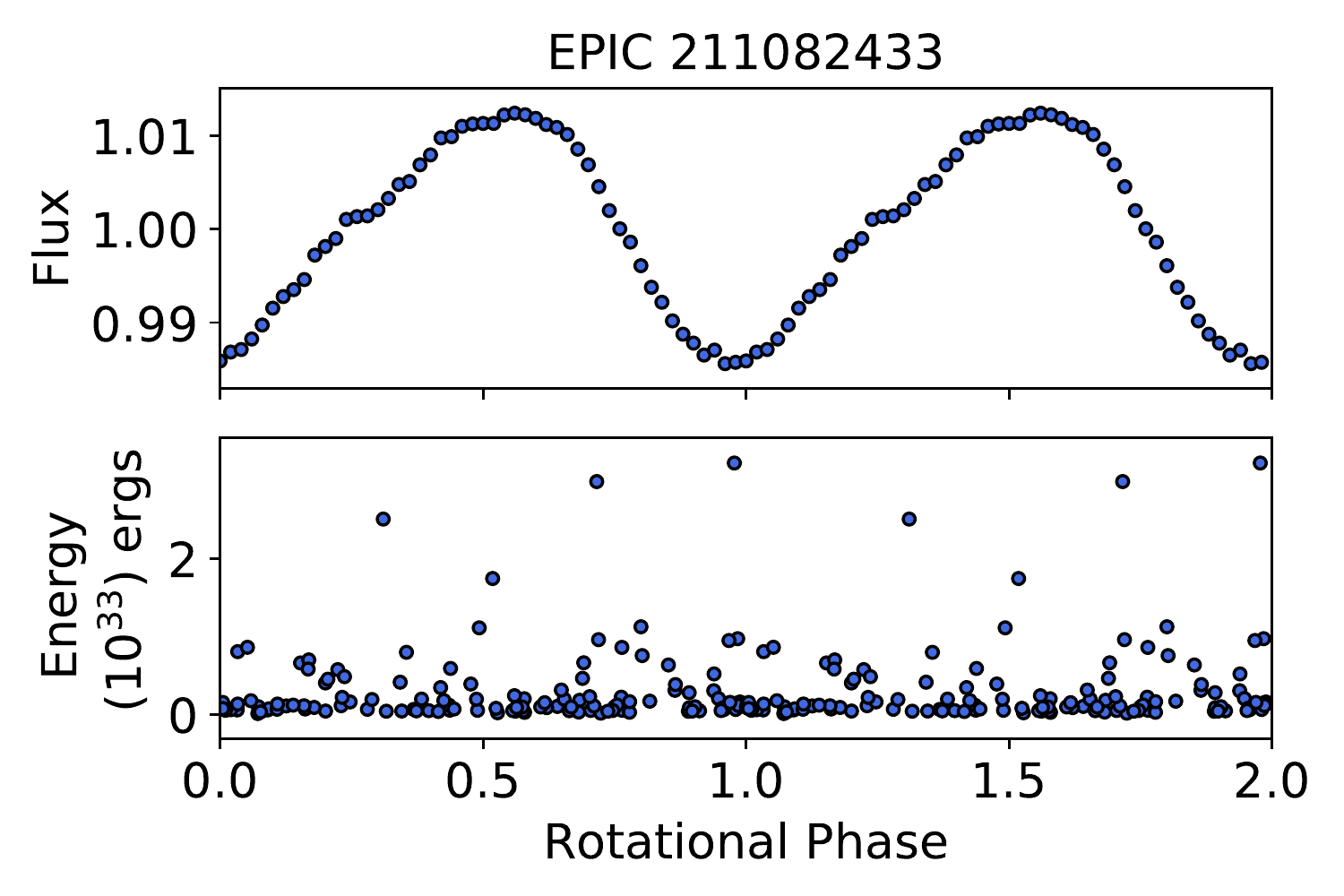}}\\
\subfloat[]{\includegraphics[width=.35\textwidth]{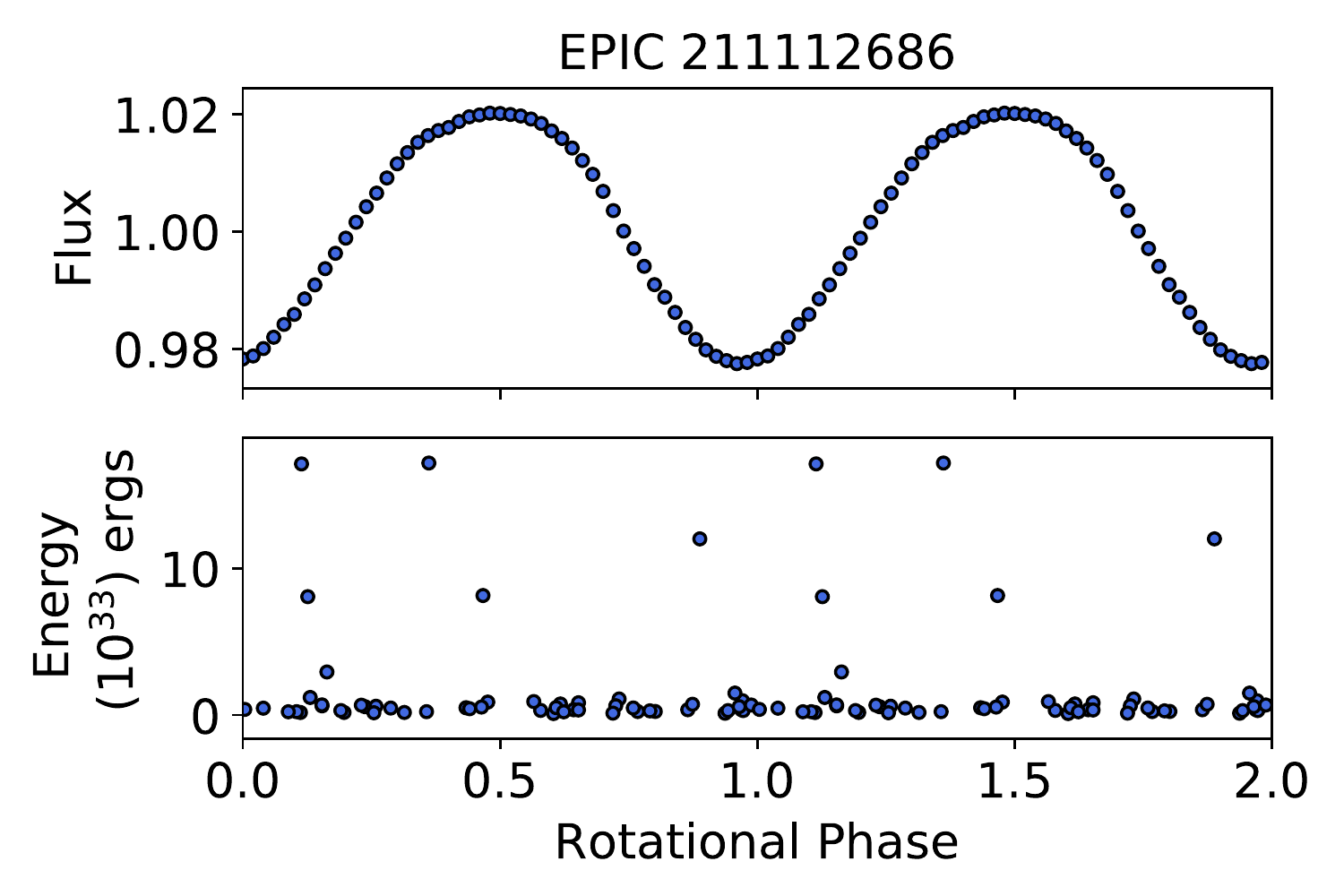}}
\subfloat[]{\includegraphics[width=.35\textwidth]{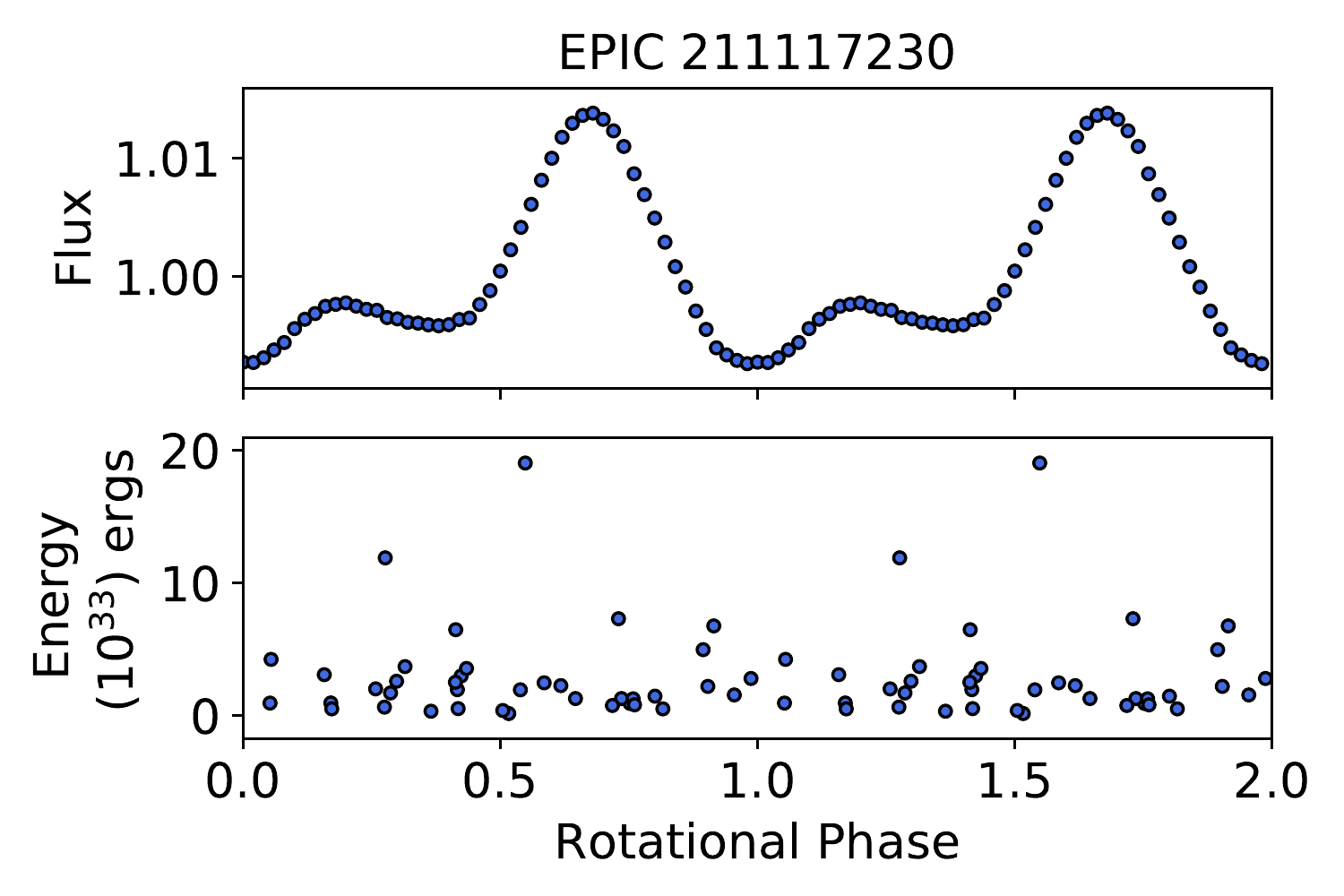}}
\subfloat[]{\includegraphics[width=.35\textwidth]{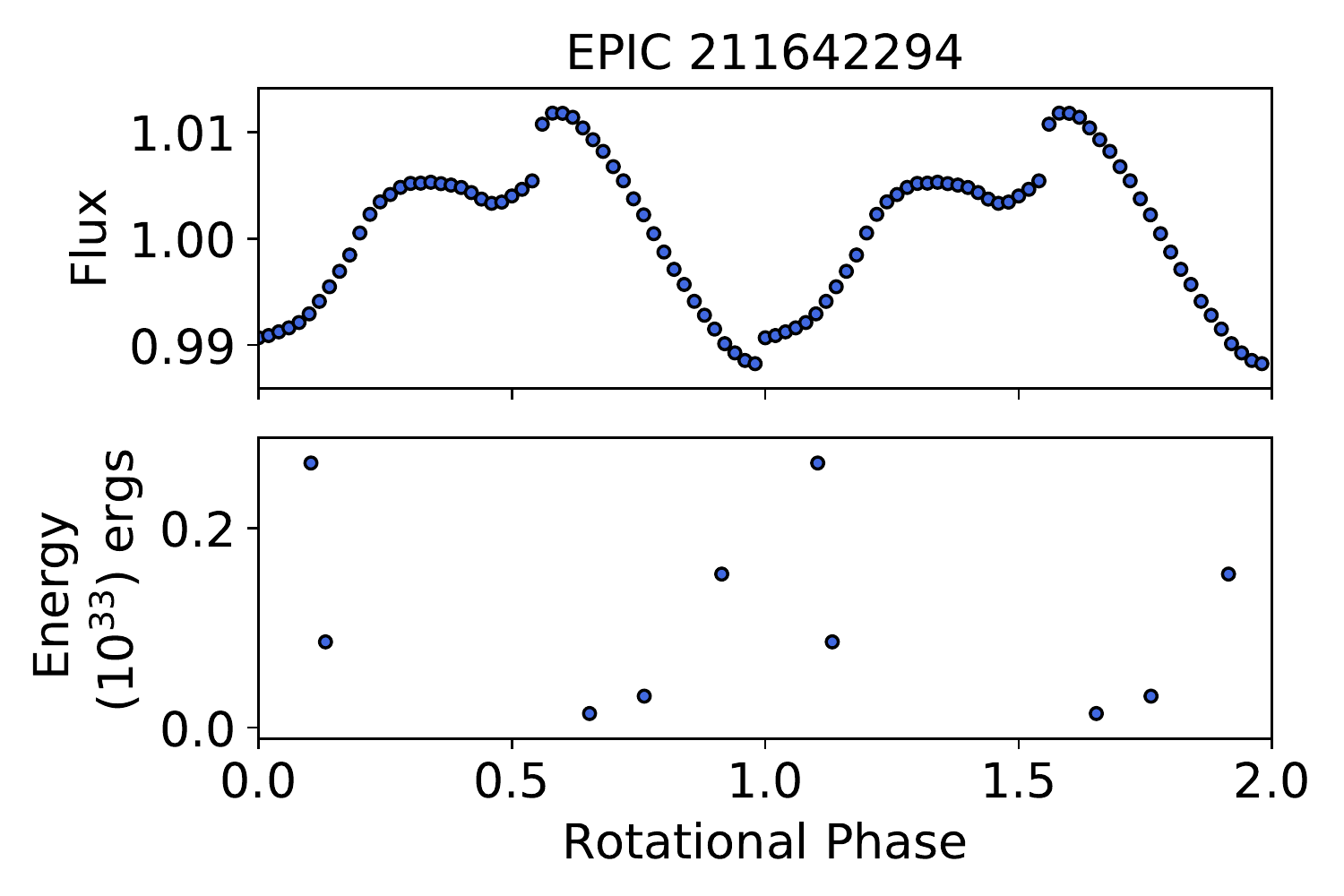}}\\
\subfloat[]{\includegraphics[width=.35\textwidth]{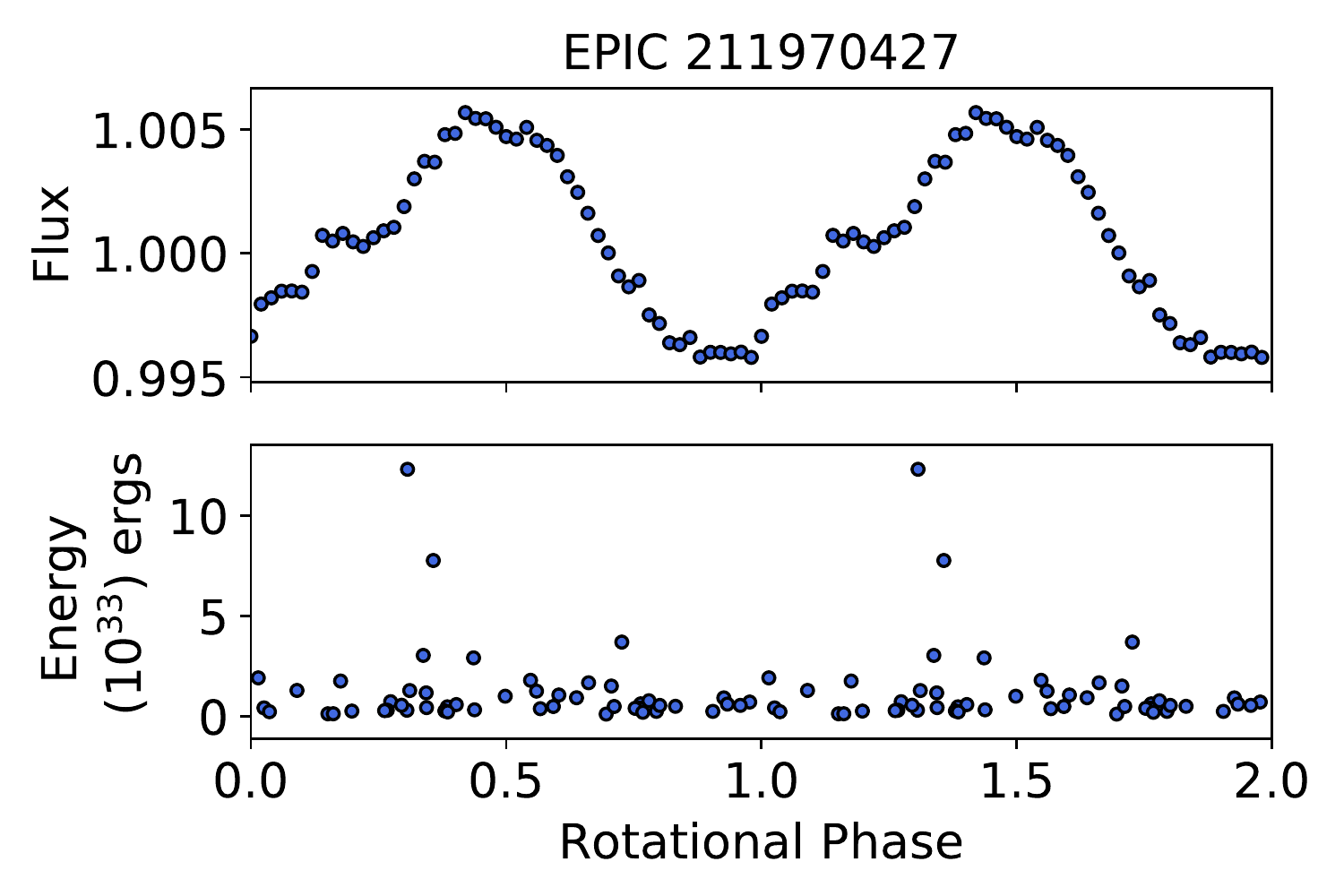}}
\subfloat[]{\includegraphics[width=.35\textwidth]{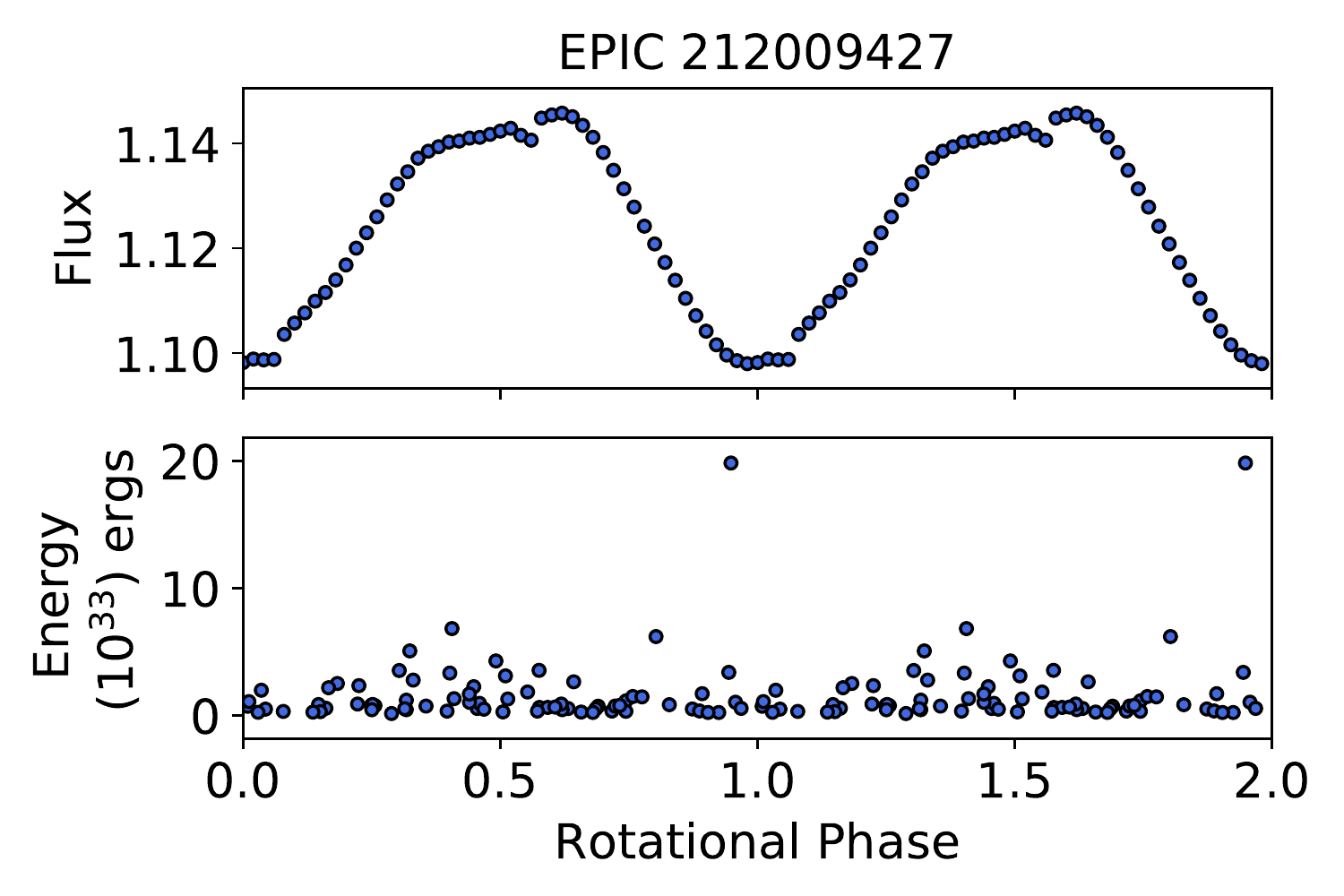}}
\subfloat[]{\includegraphics[width=.35\textwidth]{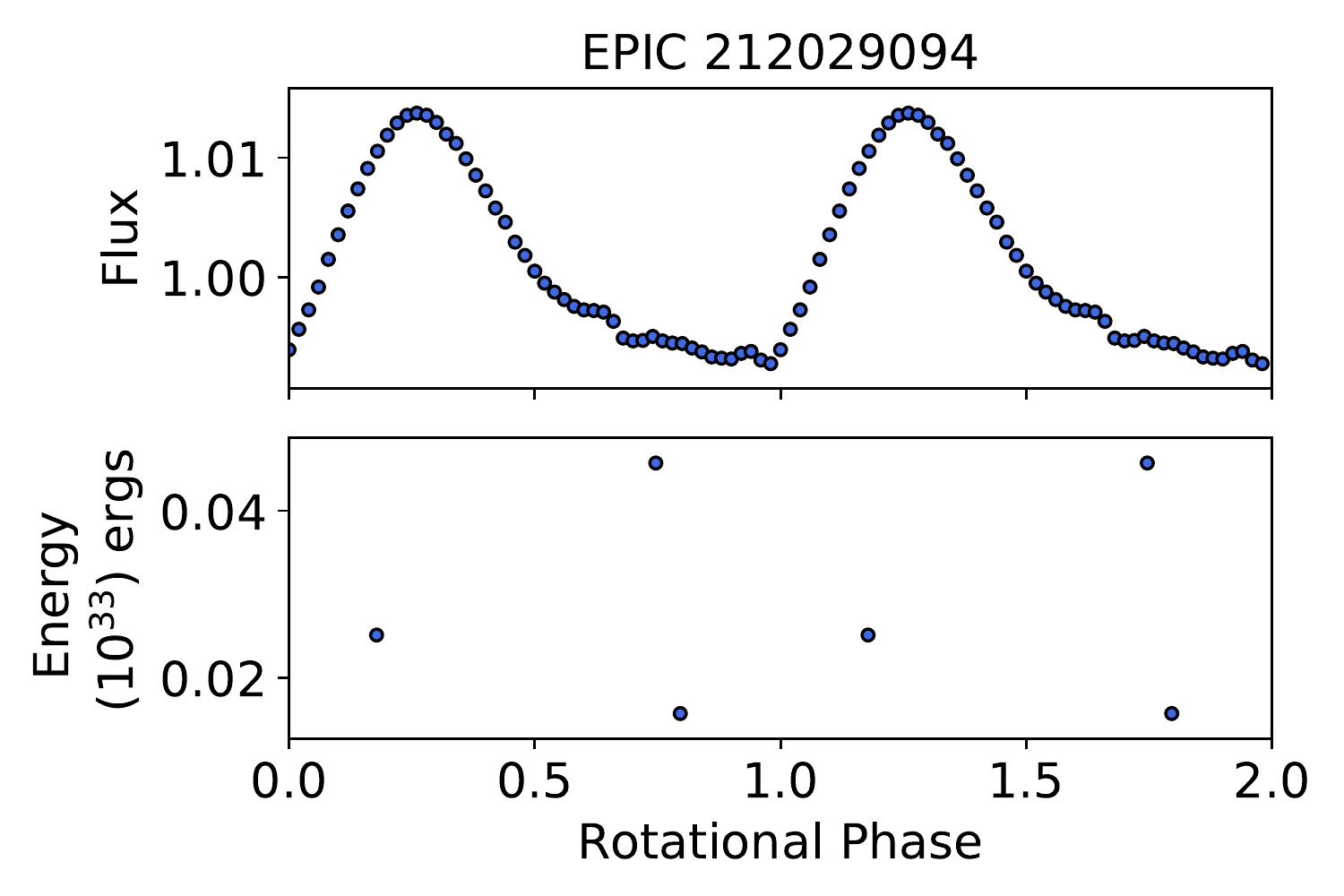}}\\
\subfloat[]{\includegraphics[width=.35\textwidth]{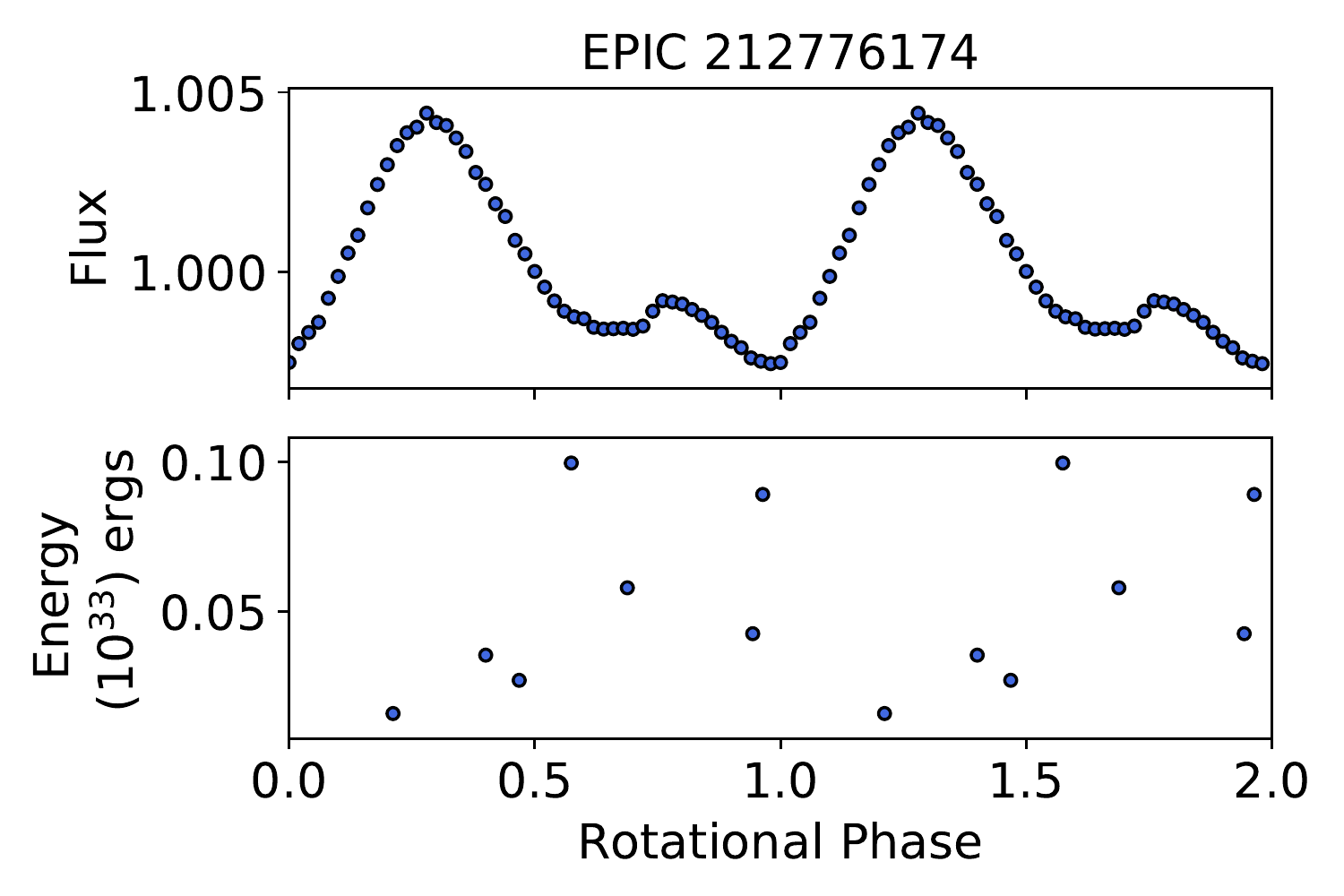}}
\subfloat[]{\includegraphics[width=.35\textwidth]{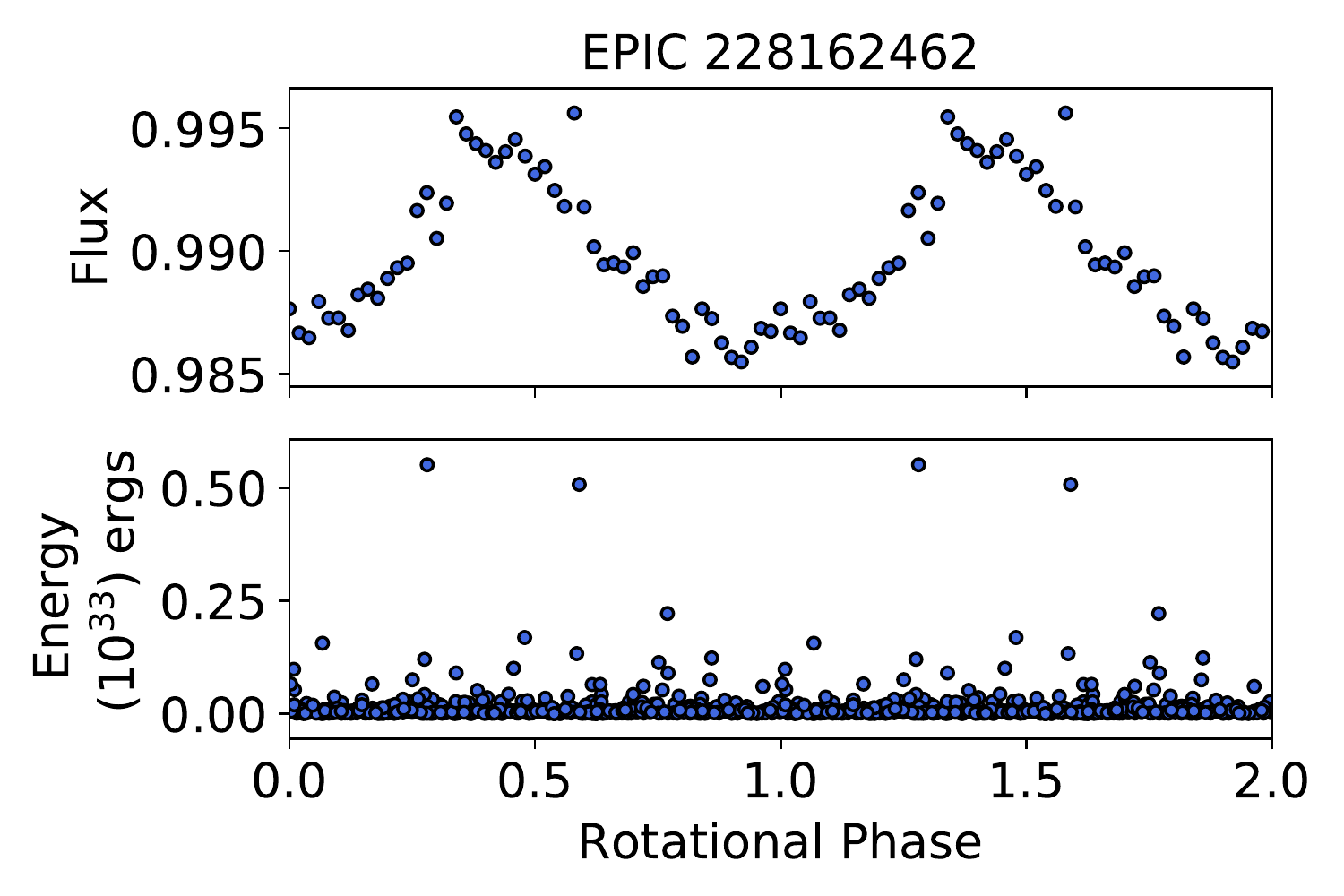}}
\caption{\textbf{(Cont.)} In the top panels we show the lightcurves of each star phased and binned on the rotation period such that there are 50 bins per rotation phase. In the bottom panels we show the phase of the flares with the energy. For each star we plot the data twice so they cover rotation phase 0.0--2.0 where 1.0--2.0 is simply a repeat of 0.0--1.0.}
\label{lightphase}
\end{figure*}

\end{document}